\documentclass[twocolumn,showpacs,preprintnumbers,amsmath,amssymb,prd,letterpaper,floatfix,nofootinbib,superscriptaddress]{revtex4}  
\usepackage{graphicx}
\usepackage{epsfig}
\usepackage{color}
\usepackage{longtable}
\usepackage{hyperref}
\usepackage{latexsym}
\usepackage{amsmath}
\usepackage{amssymb}
\usepackage{dsfont}
\usepackage{verbatim}
\usepackage{bm}

\newcommand{\I}{\mathrm{i}}  
\newcommand{\E}{\mathrm{e}}  

\newcommand{\eq}[1]{(\ref{#1})}
\newcommand{\FC}{\;,}
\newcommand{\FD}{\;.}
\newcommand{\sbar}{\overline{s}}
\newcommand{\ubar}{\overline{u}}
\newcommand{\dbar}{\overline{d}}
\newcommand{\ot}{\ensuremath{{\scriptstyle\frac{1}{2}}}}
\newcommand{\ff}[2]{\ensuremath{{\scriptstyle\frac{#1}{#2}}}}
\newcommand{\sqff}[2]{\ensuremath{\sqrt{{\scriptstyle\frac{#1}{#2}}}}}

\newcommand{\be}{\begin{equation}}
\newcommand{\ee}{\end{equation}}

\begin{document}


\date{\today}
\title{$K\,\pi$  scattering for isospin $\frac{1}{2}$ 
and $\frac{3}{2}$ in lattice QCD}

\vspace{0.2cm}

\author{C. B. Lang}
\email{christian.lang@uni-graz.at}
\affiliation{Institut f\"ur Physik, FB Theoretische Physik, Universit\"at
Graz, A--8010 Graz, Austria}

\author{Luka Leskovec}
\email{luka.leskovec@ijs.si}
\affiliation{Jozef Stefan Institute, 1000 Ljubljana, Slovenia}

\author{Daniel Mohler}
\email{mohler@triumf.ca}
\affiliation{TRIUMF, 4004 Wesbrook Mall Vancouver, British Columbia V6T 2A3, Canada}

\author{Sasa Prelovsek}
\email{sasa.prelovsek@ijs.si}
\affiliation{Jozef Stefan Institute, 1000 Ljubljana, Slovenia}
\affiliation{Department of Physics, University of Ljubljana, 1000 Ljubljana, Slovenia}

\begin{abstract}
We simulate $K\pi$ scattering in $s$ wave and $p$ wave for both isospins $I=1/2,~3/2$
using quark-antiquark and meson-meson interpolating fields.  We extract the  elastic
phase shifts $\delta$ at several values of the $K\pi$ relative momenta. The resulting phases
exhibit qualitative agreement with the experimental phases in all four channels.   We
express the   $s$ wave  phase shifts near threshold in terms of the scattering length and
the effective range. Our $K\pi$  system has zero total momentum  and is simulated  on a
single ensemble with  two dynamical quarks, so results  apply for $m_\pi \simeq 266~$MeV
and $m_K\simeq552~$MeV  in our simulation. The backtracking contractions in both $I=1/2$
channels are handled by the use of Laplacian-Heavyside smeared quarks within the  distillation
method. Elastic phases are extracted from the energy levels using L\" uscher's
relations. In all four channels we observe the expected $K(n)\pi(-n)$ scattering
states, which are shifted due to the interaction. In both attractive $I=1/2$ channels we
observe additional  states that are related to resonances; we attribute them to  
$K_0^*(1430)$ in $s$ wave and $K^*(892)$, $K^*(1410)$ and $K^*(1680)$ in $p$ wave. 
\end{abstract}

\maketitle 


\section{Introduction}

Lattice QCD (LQCD) provides an approach for {\it{ab initio}} calculations of hadron
properties.  Only in recent years, the tools are becoming efficient enough to
study strong interactions between hadrons and strong decays of hadronic resonances. 
Due to the finite spatial volume, the spectral density of
scattering processes is intrinsically discrete, and thus one has to infer
scattering information from the observed energy levels. In current 
simulations these levels are scarce, and with available methods, phase shifts $\delta$ can only be determined at a limited number of 
values of the invariant mass $\sqrt{s}$. Given these limitations, it is promising that
recently several lattice studies  \cite{Lang:2011mn,Gockeler:2008kc,Feng:2010es,Feng:2011ah,Frison:2010ws,Aoki:2011yj,Pelissier:2011ib} determined the   $\pi\pi$   scattering phase shift  in $p$ wave with  $I=1$, where the
$\rho$ meson dominates. The related  $\pi\pi$ scattering in $s$ wave with $I=2$, which does not require backtracking contractions, has   been thoroughly explored on the lattice recently \cite{Dudek:2010ew,Beane:2011sc,Dudek:2012gj}.

Continuing along that path, we study here the $K\pi$ system in $s$ wave and $p$ wave,
for both, $I=1/2$ and the exotic $I=3/2$ channels. Whereas the $I=1/2$
$p$ wave is dominated by the well established and narrow resonance $K^*$, the
experimental $I=1/2$ $s$-wave phase shows a very broad  structure without clear
resonance signal below $1~$GeV. The discussion whether this should be interpreted
as a wide resonance $K_0^*(800)$  (also called $\kappa$) with a width almost as
large as its mass is continuing. It would fit into  the $0^+$ multiplet of
scalar mesons (together with the partner states $f_0$, also called $\sigma$, and $a_0$).

\subsection{Experiments}\label{sec:experiments}

The experiments provide the magnitude and phase of the scattering amplitude\footnote{Our
$T_\ell$ is denoted by $a_\ell/\sqrt{2\ell+1}$ in experiments
\cite{Estabrooks:1977xe,Aston:1987ir}.} $T_\ell^{I}=|T_\ell^I|\E^{\I\phi_\ell^I}$ as a function of the
$K\pi$ invariant mass $\sqrt{s}$. In the elastic region, both   $|T_\ell^I|$
and  $\phi_\ell^I= \delta_\ell^I$ are related to the elastic scattering phase $\delta_{l}$
\begin{equation}
\label{T}
T_\ell^{I}=\sin \delta_\ell^I~\E^{\I\delta_\ell^I}=\frac{\E^{2i\delta_\ell^I}-1}{2i}~.
\end{equation}

In early experiments, $K\pi$ scattering amplitudes were derived from analysis of  $K p\to
K\pi n$ and $Kp \to K\pi\Delta$ 
\cite{Mercer:1971kn,Bingham:1972vy,Linglin:1973xs,Matison:1974sm,Estabrooks:1977xe,Aston:1987ir}.  
Even today, the most accurate data on scattering amplitudes  is based on Estabrooks 
 \cite{Estabrooks:1977xe}  and Aston  \cite{Aston:1987ir}, and we use these
two for comparison with our lattice results. Estabrooks \cite{Estabrooks:1977xe} 
measured  both processes and was the only experiment that was able to disentangle $I=1/2$
and $I=3/2$ channels. Their phases $\delta_{\ell=0,1}^{I=1/2,3/2}$ are plotted in Fig.~\ref{fig:phases}, and all the presented  points from Ref. \cite{Estabrooks:1977xe}   are in the
elastic region.   

Aston \cite{Aston:1987ir} considered only $K p\to K\pi n$, so they were able to provide
only the sum $T_\ell=T_\ell^{1/2}+\tfrac{1}{2} T_\ell^{3/2}=|T_\ell |\E^{\I \phi_\ell}$. The phases
$\phi^{3/2}_{\ell\geq 1}$ are believed to be small (which was explicitly
confirmed by Estabrooks),  so   we  compare to their   $\delta_{\ell=1}^{I=1/2}\simeq
\phi_{\ell=1}$ in Fig.~\ref{fig:phases}.  We also compare to their  $\delta_{\ell=0}^{I=1/2}$,
which is obtained taking their $T_{0}$ \cite{Aston:1987ir} and 
subtracting\footnote{Instead of subtracting the measured $T^{3/2}_{0}$
\cite{Estabrooks:1977xe} we  subtract the effective-range formula fit through this data.
} $T^{I=3/2}_{0}$ \cite{Estabrooks:1977xe}.  The (blue) stars in Fig.~\ref{fig:phases}
correspond to Aston's phases which are fully in the elastic region, i.e., the amplitudes lie
on the unitary Argand circle with radius $|2T_\ell^I-\I|=1$. The (green) crosses  present
measured phases $\delta_\ell^I=\Phi_\ell^I$  from the same experiment  for which the amplitudes
are almost elastic, i.e., the radius of the Argand circle is  allowed to be in the range
$0.85 <|2T_\ell^I-\I|<1.15$.

The kinematics of $K\,\pi$-scattering, the analytical structure of the partial wave amplitudes and  the experimental results until the late 1970s have been reviewed in Ref. \cite{Lang:1978fk}.  

Newer results on $K\pi$ scattering were derived from  $D\to K\pi\pi$ \cite{Aitala:2005yh,Link:2009ng,Magalhaes:2011sh} and $B\to D^* K$ and sequential $D^*$ decay \cite{Poluektov:2004mf,Abe:2005ct,Aubert:2008bd,Aubert:2008zu}, but none of these studies performs the isospin decomposition of the amplitudes. 

\subsection{Theory: Continuum}


 In particular the existence and parameters of the scalar $K_0^*(800)$ and its
partner scalar nonet states have been a continuing source of discussion.

Continuum calculations have been based on  unitarized quark models
\cite{vanBeveren:1986ea,vanBeveren:2006ua,Tornqvist:1995kr,Close:2002zu,Zhou:2010ra},  chiral perturbation theory (ChPT)
\cite{Dobado:1996ps,Ananthanarayan:2000cp} and unitarized ChPT
\cite{Oller:1998hw,Oller:1998zr,Oller:2000ma,Jamin:2000wn,GomezNicola:2001as,Pelaez:2004xp,Guo:2011pa,Nebreda:2010wv}.

Unitarized ChPT expansions have also been used to study
finite volume effects \cite{Oset:2011ce,Doring:2011nd,Doring:2012eu,Bernard:2010fp} 
in order to understand what features to expect from lattice calculations, particularly for the scalar channel.  

The scattering amplitudes were parametrized 
in the most general way allowed by quantum field theory in Refs. \cite{Buettiker:2003pp,DescotesGenon:2006uk}  according to the Roy-Steiner approach; the position of the $K_0^*(800)$ and $K^*(892)$ poles in the complex plane were then derived by 
using the experimental knowledge of $K\pi$ scattering amplitudes  at high $\sqrt{s}$ \cite{Estabrooks:1977xe,Aston:1987ir}.

Examples of further analytical studies related to the $K_0^*(800)$ are given in Refs. \cite{Black:1998zc,Black:1998wt,Black:2000qq} and references therein.







\subsection{Theory: Previous lattice studies}


Up to now, lattice simulations of $K\pi$ scattering  have extracted only the  $s$-wave phase shifts $\delta_{\ell=0}^{I=3/2,1/2}$ close to the threshold, which are commonly expressed in  terms of the scattering lengths $a_0^{I=1/2,3/2}$, defined in Eq. (\ref{effective_range}).  Lattice simulations have not yet extracted the $K\pi$ $s$-wave phase shifts away from threshold and that is one of the purposes of the present work. Simulations have also not considered $p$ wave $K\pi$ phase shifts, and we aim to determine them here.  

The extraction of the  $s$-wave phase shifts near the threshold  was mainly focused on the $I=3/2$ 
channel $K^+\pi^+$, since it does not require the evaluation of challenging backtracking 
contractions. The scattering length was determined from the finite volume energy shift $\Delta E=E-m_
\pi-m_K$  utilizing  L\"uscher's formula \cite{Luscher:1986pf}.   The quenched simulations \cite{Miao:2004gy,Nagata:2008wk} were followed by NPLQCD using 2+1 staggered sea quarks and domain-wall valence quarks \cite{Beane:2006gj},  by PACS-CS using 2+1 Wilson sea and valence quarks on 
$V=32^3
\times 64$ \cite{Sasaki:2009cz} and by Fu using 2+1 staggered sea and valence quarks \cite{Fu:2011wc}. We compare all results as a function of $m_\pi$ in Fig.~\ref{fig:a_scat} by showing the 
ratio $a_0/\mu_{K\pi}$ with the reduced mass $\mu_{K\pi}=m_\pi m_K/(m_\pi+m_K)$, since the quantity 
$a_0/\mu_{K\pi}$ is independent of $m_{\pi,K}$ in lowest-order ChPT.  
ChPT  \cite{Sasaki:2009cz,Fu:2011wc} or mixed ChPT \cite{Beane:2006gj} is used to extrapolate  the 
results
derived at higher pion masses down to the physical point. The ChPT expansions
for the $K\pi$ system  are considered in Refs. \cite{Weinberg:1966kf,Griffith:1969ph,Bernard:1990kw,Roessl:1999iu,Bijnens:2004bu,Schweizer:2005nn}.

The $I=1/2$, $l=0$ channel involves also challenging  backtracking contractions. The  
scattering length $a^{I=1/2}_0$ was determined in the quenched simulation \cite{Nagata:2008wk}, then 
in the dynamical studies by PACS-CS \cite{Sasaki:2009cz} and Fu \cite{Fu:2011wc},  mentioned 
already above. The results  are compiled in Fig~\ref{fig:a_scat}. 
NPLQCD \cite{Beane:2006gj} extracted $a^{I=1/2}_0$ in the chiral limit only indirectly through the knowledge of low energy 
constants, without actually simulating the $I=1/2$ contractions. 

The  extraction of the phase shift $\delta_{0}^{1/2}$ from the first excited energy state was actually 
done in a simulation with dynamical staggered fermions in Ref. \cite{Fu:2011xb}. However, note that 
the ground state in such a simulation corresponds to the staggered taste $K_5\pi_5$, while different unphysical tastes $K_b\pi_b$ with $b\ne 5$ 
\cite{Prelovsek:2005rf,Bernard:2007qf,Bernard:2006gj}\footnote{The $K\pi$ interpolator can be 
projected to desired taste of both mesons, while the $\bar su$ interpolators used in \cite{Fu:2011xb} 
inevitably couple to all tastes as shown in Refs. \cite{Prelovsek:2005rf,Bernard:2007qf,Bernard:2006gj}, so the variational analysis is expected to render $K_b\pi_b$ as excited 
states. } are expected to contribute to excited states. Therefore, the phase shifts extracted from the excited states in such a simulation may correspond 
merely to staggered artifacts rather than  physics of $K\pi$ scattering.

The indirect lattice determination of the $K\pi$ $s$-wave phase shifts in the $I=1/2$ channel was addressed through the simulations of the scalar semileptonic $K\to \pi$ form factor $f_0$ in Ref. \cite{Flynn}. 

To summarize, there has been no direct simulation of  the $s$ wave $K\pi$ phase shifts away from threshold, and no simulation of the  $p$ wave $K\pi$ phase shifts. These are addressed in the present work.  

\section{Analysis tools}
\subsection{Energy levels and phase shift}

In LQCD one determines Euclidean correlation functions, in the simplest case
those of products of two interpolators $O_{i,j}$ with the quantum numbers of the hadronic
channel at some Euclidean time distance. For finite volumes the spectral
function  is no longer continuous and thus all information has to be derived
from the discrete energy levels. These are hidden in the spectral representation
of the correlation function for a set of interpolators $O(t)$ with the same quantum numbers 
\be
C_{jk}(t)=\langle O_j(t) O_k(0)^\dagger\rangle=
\sum_n \langle O_j(0)|n\rangle \E^{-E_n t}\langle n| O_k(0)^\dagger\rangle\FD
\ee
The state-of-the-art method to recover the low lying energy levels $E_n$ is the variational method \cite{Michael:1985ne,Luscher:1985dn,Luscher:1990ck,Blossier:2009kd}. 
The generalized eigenvalue problem 
$C(t)\vec u_n(t)=\lambda_n(t)C(t_0)\vec u_n(t)$ then disentangles the
eigenstates $n$ making it possible to obtain energy levels from the exponential decay of the
eigenvalues
\begin{equation}
\lambda_n(t)\to \E^{-E_n (t-t_0)}\,,
\end{equation}
while the effective energies 
\begin{equation}
\label{Eeff}
E_n(t)=\log \frac{\lambda_n(t)}{\lambda_n(t+1)}
\end{equation}
render $E_n(t)\simeq E_n$ at large $t$. In actual calculations it is not possible to have a complete set of interpolators allowing one to represent the physical eigenstates.
One is limited to a reasonable subset. Also, the statistical quality of $C_{jk}(t)$
is an issue. The reliability of the obtained  energy levels decreases 
for higher $|n\rangle$, the ground state being the most  reliable one. 

Effective mass plots  $E_n(t)$ [see Eq. (\ref{Eeff})] are used only to estimate the fit range for the 
exponential fits to the eigenvalues. The energy values are extracted using
correlated fits of $\lambda_n(t)$ to one or two exponentials. When using two-exponential fits starting at small $t$,
we verify that the extracted levels agree with  results obtained
from one-exponential fits starting at larger $t$. 

The information on the $K\pi$ scattering is contained in the scattering amplitudes $T_\ell^I(s)$ in Eq. (\ref{T}). Here we concentrate on  projections  to isospin $I=1/2,3/2$ and partial waves $l=0,1$. 
We choose the total  3-momentum $\mathbf{P}$ of the $K\pi$ system to be zero, so the  lattice frame represents also the center-of-momentum frame in our current simulation: 
\be
\mathbf{P}=\mathbf{p}_\pi+\mathbf{p}_K=\mathbf{0}\;,
\;\; \mathbf{p}_\pi=-\mathbf{p}_K=\mathbf{p}^*\;,
\;\;  s=E^2-\mathbf{P}^2=E^2\,.
\ee
We measure the  energy $E$ of the interacting $K\pi$ system in finite volume 
\be 
\label{energy}
E=\sqrt{s}=\sqrt{(p^K+p^\pi)^2}=\sqrt{p^{*2}+m_\pi^2}+\sqrt{p^{*2}+m_K^2}\;,
\ee
(here $p^K$ and $p^\pi$ denote 4-momenta) which allows us to extract  the momentum $p^{*}=|\mathbf{p}^*|$ and the dimensionless $q$ via  
\begin{equation}\label{pstar_q2}
p^{*2}=\frac{[s-(m_K+m_\pi)^2][s-(m_K-m_\pi)^2]}{4s}~,\quad q\equiv p^* \frac{L}{2\pi} \FD
\end{equation}
Assuming that the strong interaction of $K\pi$ is localized to $r<R$, 
the extracted $p^*$ represents  the momenta of $\pi$ and $K$ in the outer region $r>R$. 
The resulting momenta $p^*$ or $q$ can be related to the scattering phase shift in the elastic region  
\cite{Luscher:1985dn,Luscher:1986pf,Luscher:1990ux,Luscher:1991cf}
\begin{equation}\label{zeta}
\tan \delta(q)=\frac{\pi^{3/2} q}{\mathcal{Z}_{00}(1;q^2)}\qquad \mathrm{for}\ \mathbf{P}=0~.
\end{equation}
This is L\" uscher's relation \cite{Luscher:1990ux} between the energy levels in finite volume [which enter via Eq. (\ref{pstar_q2})] and the phase shifts. The generalized zeta function $Z_{lm}$ is given in Ref. \cite{Luscher:1990ux}.

In the discussion of our results, we also apply the combination 
\be\label{def_rho}
\rho_\ell^I(s)\equiv\frac{(p^*)^{2\ell+1}}{\sqrt{s}} \cot\delta_\ell^I(p^*)\FC
\ee
which vanishes at the position of the resonance  $\delta(s_R)=\frac{\pi}{2}$. 
The variable $\rho_\ell^I(s)$ provides a convenient parametrization of the elastic partial wave near threshold,
\be \label{effective_range}
\sqrt{s}\,\rho_\ell^I(s)=\frac{1}{a^I_\ell} +\ot r_\ell^{I} p^{*2} +\mathcal{O}(p^{*4})\FC
\ee
where $a$ is the scattering length and $r$ the effective range. 
Near a relativistic
Breit-Wigner resonance in the elastic region we may write the partial
wave amplitude \cite{Nakamura:2010zzi} 
\begin{equation}
\label{amplitude}
T_\ell^I=\frac{-\sqrt{s}\,\Gamma(s)}{s-s_R+\I \sqrt{s}\,\Gamma(s)}=
\E^{\I\delta_\ell^I(s)}\sin \delta_\ell^I(s)\FD
\end{equation}
where $s_R=m_R^2$ denotes the resonance position and $\Gamma$ is the decay width. Considering the threshold behavior we can define  
\be \label{Gamma}
\Gamma(s)=\frac{(p^*)^{2\ell+1}}{s} ~\gamma\FC
\ee  
and get
\begin{align}\label{BW_resonance}
\sqrt{s}\,\Gamma(s)\,\cot\delta(q)&= \rho_\ell(s)~ \gamma=s_R-s\FC\nonumber\\
\rho_\ell(s)&=\frac{1}{\gamma}(s_R-s)\FD
\end{align}
For a resonance in the $l=1$ channel one has $\gamma=g^2/(6\pi)$, defining the
coupling constant $g$ (e.g., $g_{K^*K\pi}$).

Naively, one would expect that in simulations with dynamical quarks, all possible
intermediate states should contribute. In the correlation function of a
$\rho$ meson with a quark-antiquark interpolator, one thus
would expect to find signals of the $\pi\pi$ intermediate state (in $p$ wave). It
turned out that this appears not to be the case, most likely due to weak
coupling of the corresponding lattice interpolators. One had to include
$\pi\pi$ interpolators explicitly in order to obtain energy levels representing
scattering \cite{Lang:2011mn,Aoki:2011yj,Gockeler:2008kc,
Feng:2010es,Feng:2011ah,Frison:2010ws}. This motivates us to include $K\pi$ and some other 
meson-meson interpolators in  the present study.  

In  the study of the $\pi\pi$ scattering \cite{Lang:2011mn}, we also used interpolators
with total momentum $\mathbf{P}\not =0$, which allowed us to find phase shifts at more values of $s=E^2-\mathbf{P}^2$. 
The corresponding phase shift relation \eq{zeta} for scattering of particles of equal mass $m_1=m_2$ and total momentum $\mathbf{P}\not =0$ 
has been
derived in Refs. \cite{Rummukainen:1995vs,Kim:2005zzb}. For systems with two mesons
of different mass $m_1\not =m_2$ and $\mathbf{P}\not =0$, the formalism has been extended in Refs.
\cite{Fu:2011xz,Leskovec:2012gb,Doring:2012eu,Gockeler:2012yj}. 

In our present study, we consider only the case $\mathbf{P}=\mathbf{p}_\pi+\mathbf{p}_K=0$, and so the original phase shift relation \eq{zeta} applies \cite{Luscher:1990ux}. We consider the $A_1^+$ irreducible representation of $O_h$ to extract the $s$ wave and the $T_1^-$ irreducible representation  to extract the $p$ wave. $A_1^+$ will also contain the admixture of $l=4$ and higher partial waves, and $T_1^-$ will contain admixture of $l=3$ (and higher waves) \cite{Luscher:1990ux}, but those are expected to be small, and we neglect the effect of such mixing. In contrast to that, the simulations of 
 $K\pi$ scattering with $\mathbf{P}\not =0$ would involve an additional complication since 
the partial waves for even and odd $\ell$ can mix  in the same irreducible representation \cite{Fu:2011xz,Leskovec:2012gb,Doring:2012eu,Gockeler:2012yj}.  We avoid this additional
complication by taking $\mathbf{P}=0$ in the present work.

It is important to provide a large enough set of interpolators for a good
representation of the lowest physical states in the generalized eigenvalue
analysis. The interpolators used for the four channels studied (isospin
$1/2$ and $3/2$, $s$ wave and $p$ wave) are given in Appendix
\ref{app_a}. In addition to several $\overline q q$ operators ${\cal O}^{\bar qq}$ we also include
several meson-meson operators ${\cal O}^{MM}$ ($K\,\pi$, $K^*\,\rho$, $K_1 \,a_1$), 
in total up to eight for the $s$ wave and up to six for the $p$ wave. We include for example $\pi(0)K(0)$ and 
$\sum_{i=x,y,z}[\pi(\mathbf{p}_i)K(-\mathbf{p}_i)+\pi(-\mathbf{p}_i)K(\mathbf{p}_i)]$
for $s$ wave, and $\pi(\mathbf{p}_z)K(-\mathbf{p}_z)-\pi(-\mathbf{p}_z)K(\mathbf{p}_z)$ for $p$ wave ($\mathbf{p}_i=\tfrac{2\pi}{L}\mathbf{e}_i$). Here, each meson is separately projected to definite momentum, given in parentheses in units of $2\pi/L$. 
The Wick contractions are provided in Appendix \ref{app_b}. 

\subsection{Lattice simulation}

Like in Ref. \cite{Lang:2011mn} we use configurations generated for the study of
reweighting techniques \cite{Hasenfratz:2008ce,Hasenfratz:2008fg} kindly 
provided by the authors. The action used to generate
the gauge configurations containing $n_f=2$ flavors of mass-degenerate light
quarks is a tree level improved Wilson-Clover action with gauge links smeared
using one level of normalized hypercubic smearing (nHYP smearing). The practical advantage of $n_f=2$ flavor simulation  for $K\pi$ scattering 
(with respect to $n_f=2+1$) is that there is no $K\eta$ scattering state, so the inelastic threshold is higher (discussion in Sec. \ref{sec:swave_half}).    The valence $u/d$ quarks have the same mass as the sea $u/d$ quarks. 
 Table \ref{gauge_configs} lists the parameters
used for the simulation along with the number of (approximately independent)
gauge configurations used, the lattice spacing, volume and the pseudoscalar masses $m_{\pi,K}$, which are the most  relevant for $K\pi$ scattering  (for details, see Ref. \cite{Lang:2011mn}).

The $s$ quark is included only as a valence quark in the hadron propagators. To determine the strange quark hopping parameter $\kappa_s$, we calculated the connected part of the $\phi$ meson which is expected to be almost exclusively $\bar{s}s$. The tuning was done on sources in a single time slice and we obtained $\kappa_s=0.12610$. Using our complete set of perambulators calculated with this value of $\kappa_s$ we again determine the mass of the $\phi$ meson on the full data set and obtain $m_\phi^{lat}=1015.8\pm10.8~$MeV which has to be compared to the experimental mass $m_\phi^{exp}=1019.455\pm0.020~$MeV. This $\kappa_s$ corresponds to the kaon mass provided in Table \ref{gauge_configs}.

\begin{table}[t]
\begin{ruledtabular}
\begin{tabular}{ccccccc}
$N_L^3\times N_T$ & $\beta$ & $a$[fm] & $L$[fm] & \#configs & $m_\pi$[MeV]& $m_K$[MeV]\\ 
\hline
$16^3\times32$ & 7.1 & 0.1239(13) & 1.98 & 280 & 266(3)(3) & 552(2)(6)\\
\end{tabular}
\end{ruledtabular}
\caption{\label{gauge_configs} Configurations used for the current study have two light dynamical flavors. $N_L$
and $N_T$ denote the number of lattice points in spatial and time directions.
For more details see  \cite{Lang:2011mn}.}
\end{table}

The sea and valence quarks obey periodic boundary conditions in space.
The gauge field obeys periodic boundary conditions, while  the sea quarks satisfy antiperiodic 
boundary conditions in time. We compute and
combine valence quark propagators with both antiperiodic and  periodic boundary
conditions. This effectively extends the time direction to $2N_T=64$ by
combining the periodic propagator $M^{-1}_P$ and antiperiodic propagator
$M^{-1}_A$ (see for example \cite{Sasaki:2001nf,Detmold:2008yn}). All results in
this paper have been obtained using the so-called ``P+A'' propagators
\begin{align}
\label{P+A}
M^{-1}_{P+A}(t_f,t_i)=
\begin{cases}
\tfrac{1}{2}[M^{-1}_P(t_f,t_i) + M^{-1}_A(t_f,t_i)] & t_f\geq t_i\FC\\
\tfrac{1}{2}[M^{-1}_P(t_f,t_i) - M^{-1}_A(t_f,t_i)] & t_f< t_i \FD
\end{cases}
\end{align}
For further discussion of this method, see Ref. \cite{Lang:2011mn}.

\subsection{Propagators, contractions and distillation method}

The hadron correlation functions are constructed by combining quark propagators
derived on the gauge field configurations. For the meson-meson interpolators
used here, the Wick contractions lead also to backtracking contributions, for example depicted in the box diagram of Fig.~\ref{fig:contractions}c. 
For a statistically reliable inclusion one thus needs all-to-all
methods. The distillation method proposed in Ref. \cite{Peardon:2009gh} provides these
capabilities. It is based on separable quark smearing operators, i.e., a
truncated spectral representation of the unit operator in terms of the
eigenvectors  of  the spatial lattice Laplacian. The quarks $q_s$ in the simulation are smeared according to different smearing widths $s$  \cite{Lang:2011mn}
\begin{align}\label{smearing}
q_s&\equiv \sum_{k=1}^{N_v} v^{(k)} v^{(k)\dagger }~q\\
N_v&=96,\;64,\; \mathrm{or}\;32 \nonumber\\
\mathrm{for}\;s&=n\ \mathrm{(narrow)}\FC
    m\ \mathrm{(middle)}\FC \; \mathrm{or}\; w\ \mathrm{(wide)}\FC \nonumber
\end{align}
rendered by the different number $N_v$ of lowest eigenvectors $v^{(k)}$ incorporated in the sum. 
 The technique effectively 
replaces the  quark propagators $G(x\to y)$ by propagation from one source to
another, so-called perambulators $\tau(i\to j)$. This allows high flexibility in
the Dirac and momentum structure of the hadron interpolators.

We proceed as follows. First, the gauge links are four-dimensional  normalized hypercubic smeared
\cite{Hasenfratz:2007rf} with the same parameters used for generating the gauge
configurations: $(\alpha_1,\alpha_2,\alpha_3)=(0.75,0.6,0.3)$.  On each gauge
configuration, we calculate the lowest 96 eigenvectors of the lattice Laplacian
on every time slice. For the calculation of the eigenmodes, we use the PRIMME
package \cite{Stathopoulos:2009:PPI}. For the determination of the quark
propagators we use the \verb+dfl_sap_gcr+ algorithm provided in L\"uscher's
DD-HMC package \cite{Luscher:2007se,Luscher:2007es}. Due to the large number of
sources necessary for the distillation approach, an inverter employing low-mode
deflation techniques is especially well-suited.

Statistical errors are determined with a single elimination jackknife procedure
throughout. When extracting energy levels, we properly account for correlation in
Euclidean time $t$ by estimating the full covariance matrix in the given fit
interval. For the covariance matrix, we use a jackknife estimate which is
calculated on the ensemble average  only.


\section{Results}

\subsection{Pion, kaon and dispersion relations}

Since we consider scattering of $\pi$ and $K$, we need their masses and their separate (noninteracting) energies $E_{\pi,K}(p)$.  For $\pi$ and $K$ ($J^{P}=0^{-}$)  we use the six
interpolators given in Eq. \eq{pion_and_kaon_interpolators}, with three smearing
widths for each of the two Dirac structures. Their masses and energies    are extracted
from the variational analysis of the $6\times 6$ correlation matrix and listed in Table \ref{tab:dr}. 

We find that  the lattice energies $E_{\pi,K}(\mathbf{p}\!=\!\tfrac{2\pi}{L}\mathbf{n})$ agree for  $\mathbf{n}^2\leq 2$ with the continuum dispersion relation within the error (see Table \ref{tab:dr} \footnote{Table \ref{tab:dr}  indicates that  the measured energies $E$ agree better with the prediction of the continuum dispersion relation $E_{cont}^{d.r.}$ than with the prediction $E_{lat}^{d.r.}$ based on free lattice theory. This may be
due to the smearing of link and quark operators which improves rotational
invariance properties.}).  
So we use the  dispersion relation  $E_{\pi,K}(p^*)=\sqrt{m_{\pi,K}^2+p^{*2}}$ with $m_{\pi,K}$ fixed to $a\,m_\pi=0.1673$ and $a\,m_K=0.3466$  throughout the analysis. This means that $p^{*2}$ and $q^2$ are extracted from the lattice energy $E$ using 
Eqs. (\ref{energy},\ref{pstar_q2}), and the resulting $q^2$ is used to get the 
phase shift from Eq. (\ref{zeta}). 

As a cross-check, we have verified that  our resulting phase shifts obtained 
in this way agree with the phase shifts  obtained from the energy shifts $\Delta E_n=E_n-E_\pi-E_K$ of interacting $K\pi$  with respect to the nearest noninteracting scattering level. These  values are determined using the ratio $\lambda_n(t)/[\lambda_{\pi(\mathbf{n}_\pi)}(t)~ \lambda_{K(\mathbf{n}_K)}(t)]\propto \E^{-\Delta E_n t}$ of the eigenvalues $\lambda_n(t)$ for interacting $K\pi$ and the eigenvalues for noninteracting $K(\mathbf{n}_K)$ and $\pi(\mathbf{n}_\pi)$ with momenta $\tfrac{2\pi}{L}\mathbf{n}_{K,\pi}$. 

\begin{table*}[t]
\begin{ruledtabular}
\begin{tabular}{c c c c  c c c c}
$\mathbf{n}=\mathbf{p} \tfrac{L}{2\pi}$ & $t_0$ &interpol. & fit range &  $\chi^2$/d.o.f. & $E\,a$ (simul.)  & $E_{cont}^{d.r.}\,a$ & $E_{lat}^{d.r.}\,a$\vspace{3pt}\\
\hline 
(0,0,0) & 3 & ${\cal O}_{1,2}^w{\cal O}_{1,2}^m{\cal O}_{1,2}^n$  & 8-14  & 1.57/5 &$a\,m_\pi=$ 0.1673(16)  & -- &-- \\
(0,0,1) & 3 & ${\cal O}_{2}^w{\cal O}_{2}^n$      & 12-17 & 0.98/4          & 0.4374(64) & 0.4268(65) & 0.4215(65) \\
(1,1,0) & 4 &  ${\cal O}_{2}^w{\cal O}_{1}^n$     & 8-13  & 1.31/4          & 0.5823(46) & 0.5800(48) & 0.5690(47) \\
\hline 
(0,0,0) & 4 & ${\cal O}_{1,2}^n$                  & 7-16  & 10.1/8 & $a\,m_K=$ 0.34660(86)  & -- &-- \\
(0,0,1) & 4 & ${\cal O}_{1}^w{\cal O}_{2}^n$      & 9-16  & 4.22/6          & 0.5236(11) & 0.52376(58) & 0.51724(58) \\
(1,1,0) & 4 &  ${\cal O}_{1}^w{\cal O}_{2}^n$     & 8-14  & 0.89/5          & 0.6516(24) & 0.65463(46) & 0.64148(45)  \\
\end{tabular}
\end{ruledtabular}
\caption{The ground state pion and kaon energies  extracted for three momenta:  $E$ is
extracted from the variational analysis using the chosen interpolator sets. 
They are compared to analytic expectations in the continuum $E_{cont}^{d.r.}=\sqrt{m^2+p^2}$ and in the free lattice theory 
$E_{lat}^{d.r.}a=\cosh^{-1}[\cosh(ma) + 2\sum_i \sin^2(\tfrac{1}{2}p_ia)]$ where the error comes solely from $a\,m_{\pi ,K}$ above. }\label{tab:dr}
\end{table*}

\subsection{$K\pi$ scattering} 

\begin{figure*}[t]
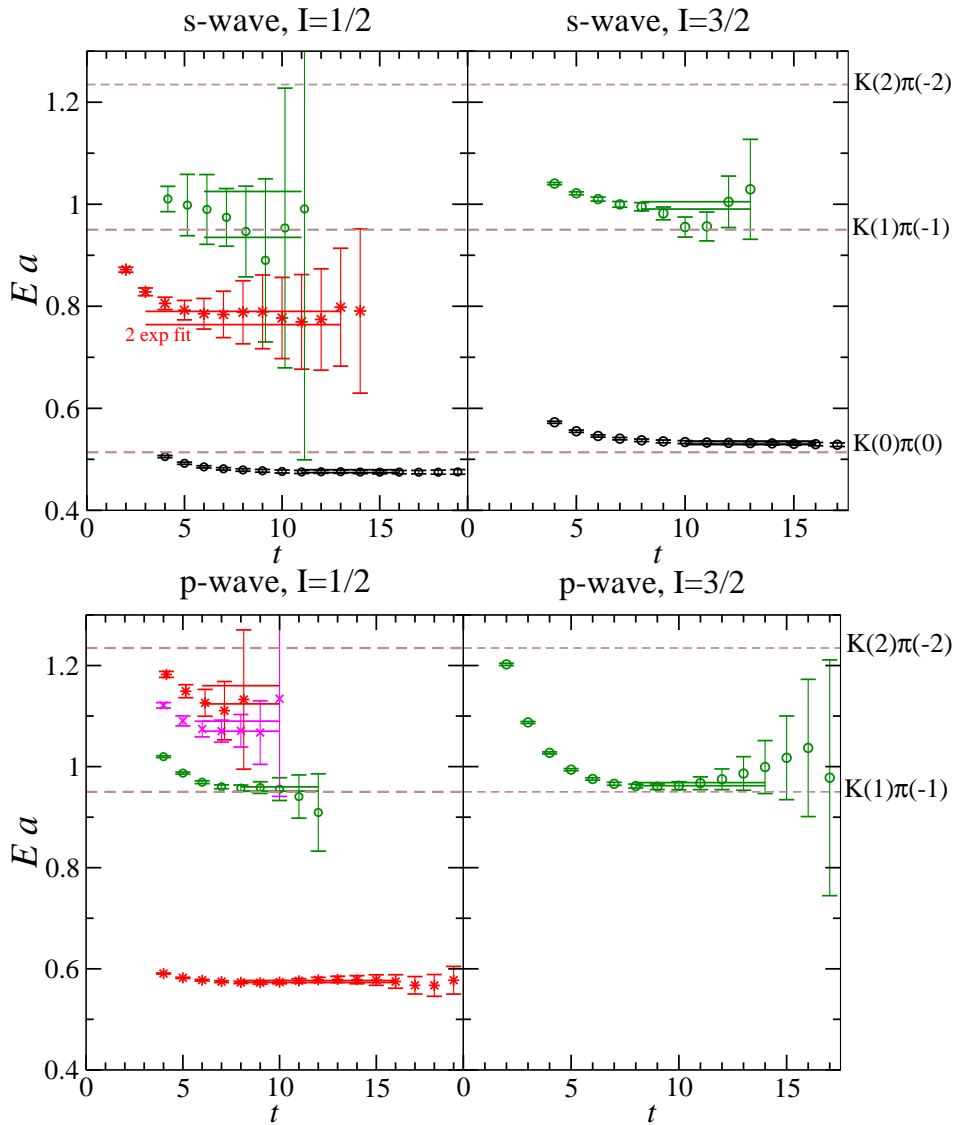

\begin{center}
\includegraphics*[width=0.7\textwidth,clip]{figs/eff_eig_s_wave_chosen.eps}
\includegraphics*[width=0.7\textwidth,clip]{figs/eff_eig_p_wave_chosen.eps}
\end{center}
\caption{
Effective energies $E(t)a$ of the lowest eigenvalues for the interpolator choices listed in Table \ref{tab:phases}, together with the resulting energies obtained with 1-exponential or 2-exponential fits. 
    The horizontal broken lines show the energies $E=E_K+E_\pi$ of the 
noninteracting scattering states $K(n)\pi(-n)$ as measured on our lattice;  
 $K(n)\pi(-n)$  corresponds to the scattering state with $p^*=\sqrt{n}\tfrac{2\pi}{L}$. Note that   there is no $K(0)\pi(0)$  scattering state for $p$ wave. Black and green circles correspond to the shifted scattering states, while 
 the red stars and pink crosses  correspond to additional states related with resonances.}\label{fig:eff_eig_chosen}
\end{figure*}

Before addressing the details of the analysis and the results for separate channels below, 
let us compare the main features of all four channels ($s$ wave and $p$ wave in $I=1/2,~3/2$).   
The resulting energy levels for the $K\pi$ system are presented in terms of the effective energy 
$E(t)a$ [see Eq. (\ref{Eeff})] in Fig.~\ref{fig:eff_eig_chosen}. These levels correspond to the preferred interpolator choices listed in Table \ref{tab:phases}. The horizontal broken lines show the energies $E=E_K+E_\pi$ of the 
noninteracting states $K(n)\pi(-n)$ as measured on our lattice with $p^*=\tfrac{2\pi}{L}\,\sqrt{n}$. 

The resulting spectrum  agrees with the expectations for the respective channels: 
there is a scattering state $K(0)\pi(0)$ in $s$ wave (black circles), which is below $m_K+m_\pi$ in the attractive $I=1/2$ channel and above $m_K+m_\pi$ in the repulsive $I=3/2$ channel. There is no scattering state $K(0)\pi(0)$ in $p$ wave due to nonvanishing orbital momentum. The scattering state $K(1)\pi(-1)$ (green circles at $Ea\simeq 0.95$) is observed in all four channels, and its signal is nicer 
for $I=3/2$ (with $l=0,1$), since channels with maximal isospin $I=3/2$  do not involve 
backtracking contractions. In addition to the scattering states, which lie close to noninteracting levels given by the dashed lines, there are additional states in the attractive $I=1/2$ channels (red and pink  levels). While these states are of course also shifted with regard to the resonance position, we refer to these additional states as ``related to'' the respective resonance. In this language, the additional state in the $s$ wave is probably related to the scalar resonance  $K_0^*(1430)$, while the additional states in $p$ wave are expected to be related to vector resonances $K^*(892)$, $K^*(1410)$ and $K^*(1680)$, respectively.  

\begin{figure*}[t]
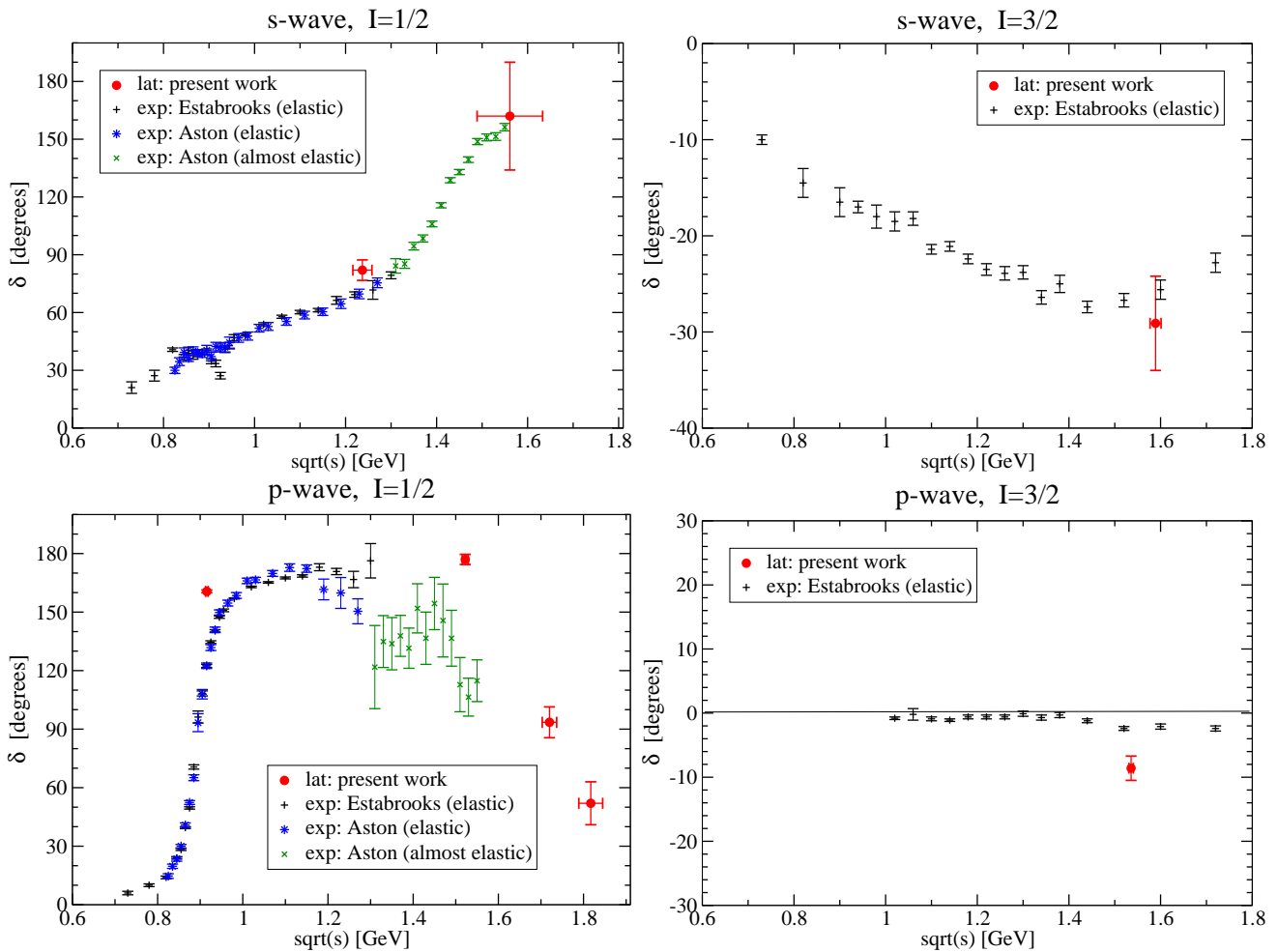

\begin{center}
\includegraphics*[width=0.48\textwidth,clip]{figs/delta_s_half_sqrts_phys.eps}
\includegraphics*[width=0.48\textwidth,clip]{figs/delta_s_threehalf_sqrts_phys.eps}
\includegraphics*[width=0.48\textwidth,clip]{figs/delta_p_half_sqrts_phys.eps}
\includegraphics*[width=0.48\textwidth,clip]{figs/delta_p_threehalf_sqrts_phys.eps}
\end{center}
\caption{
 The extracted $K\,\pi$ scattering phase shifts $\delta_\ell^I$  in all four channels $l=0,1$ and $I=1/2,~3/2$. The phase shifts are shown as a function of the $K\pi$ invariant mass $\sqrt{s}=M_{K\pi}=\sqrt{(p_\pi+p_K)^2}$. Our  results (red circles) apply for  $m_\pi\simeq 266~$MeV  and $m_K\simeq 552~$MeV in our lattice simulation. In addition to the phases provided in four plots, we also extract the values of $\delta_0^{1/2,~3/2}$  near threshold $\sqrt{s}=m_\pi +m_K$, but these are provided  in the form of the scattering length in the main text (as they are particularly sensitive to $m_{\pi ,K}$).     
Our lattice results are compared to the experimental elastic phase shifts from Estabrooks (black pluses) \cite{Estabrooks:1977xe} and Aston (blue stars) \cite{Aston:1987ir}. Dark green crosses represent measured phase shifts by Aston  \cite{Aston:1987ir} which correspond to an almost elastic amplitude $T_\ell^I$, i.e., $0.85 < |2T_\ell^I-i|<1.15$ (see Sec. \ref{sec:experiments}). Lattice phase shifts are determined up to multiples of $180^\circ$ from Eq. (\ref{zeta}).}\label{fig:phases}
\end{figure*}

The results for phase shifts $\delta_\ell^I(s)$, which are based on the energy levels of Fig.~\ref{fig:eff_eig_chosen}, are presented in Table \ref{tab:phases}. They are compared to the experimental phase shifts in Fig.~\ref{fig:phases}. The phase shifts $\delta^{1/2,3/2}_0$ for $\sqrt{s}\simeq m_\pi+m_K$ near threshold are omitted from Fig.~\ref{fig:phases} and are expressed in terms of the scattering length below. The lattice values of the phase shifts presented\footnote{Except for lowest level in $p$ wave $I=1/2$ scattering that is due to $K^*$ and will be discussed in detail below.}  in Fig.~\ref{fig:phases}  apply to $\sqrt{s}$ quite far away from the threshold and we expect that they are not significantly influenced by the exact position of the threshold, which is at $\simeq 140+500~$MeV in experiment and at $\simeq 266+552~$MeV in our lattice simulation. 

\begin{figure*}[t]
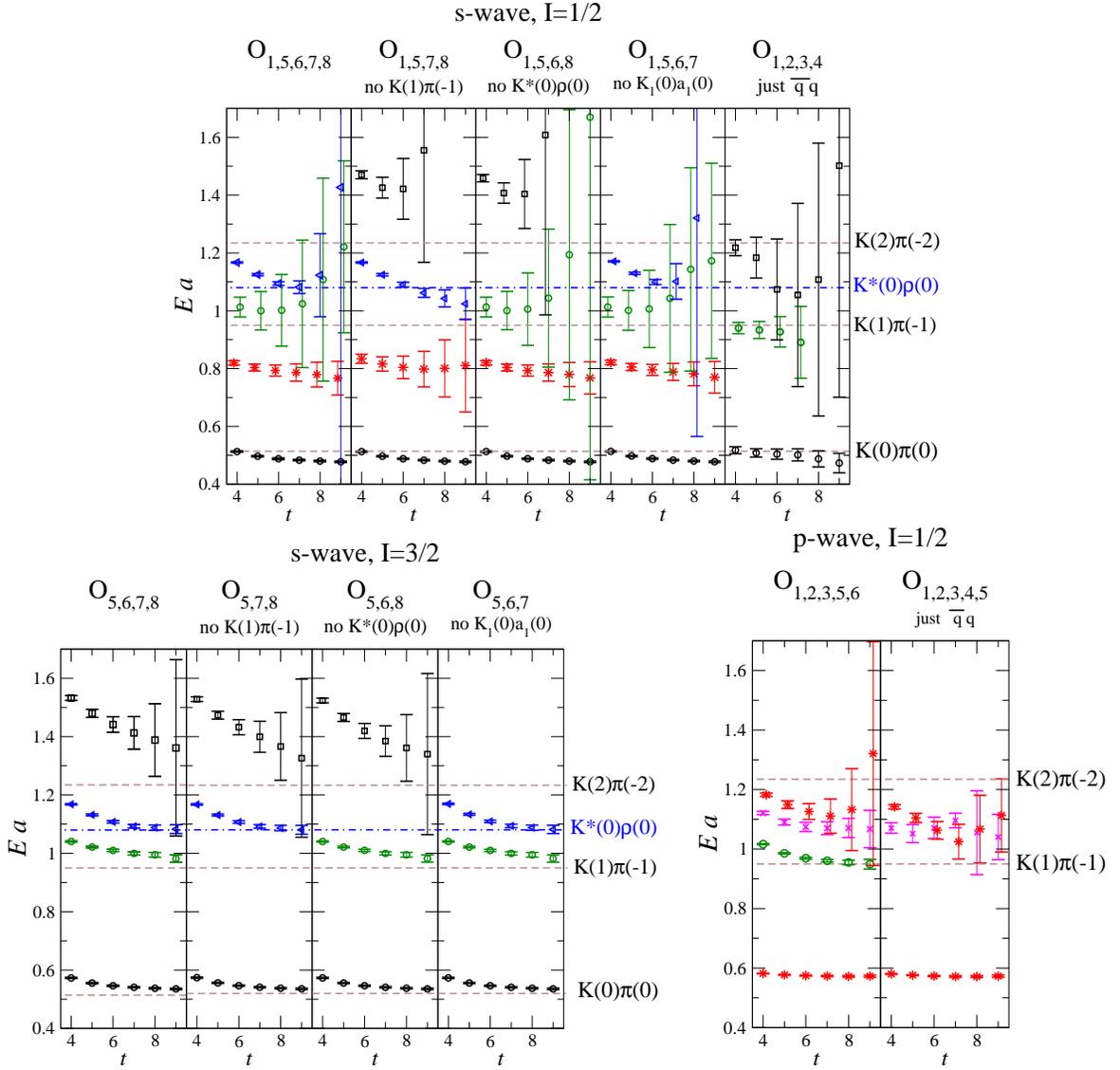

\begin{center}
\includegraphics*[width=0.6\textwidth,clip]{figs/eff_eig_s_wave_half_set_dep.eps}
\vspace{0.5cm}
\includegraphics*[width=0.51\textwidth,clip]{figs/eff_eig_s_wave_threehalf_set_dep.eps} $\quad$ 
\includegraphics*[width=0.32\textwidth,clip]{figs/eff_eig_p_wave_half_set_dep.eps}
\end{center}
\caption{
Effective energies $E(t)\,a$ of the lowest eigenvalues for different interpolator choices in the correlation matrix. The horizontal broken lines show the energies $E=E_K+E_\pi$ of the 
noninteracting scattering states $K(n)\pi(-n)$ as measured on our lattice with $p^*=\tfrac{2\pi}{L}\,\sqrt{n}$. Note that there is no $K(0)\pi(0)$ and $K^*(0)\rho(0)$ scattering state for the $p$ wave. Red stars and pink crosses  correspond to states related to resonances; other levels are related to scattering states. 
(a) $s$ wave, I=1/2:   
the first four choices incorporate also meson-meson interpolators ${\cal O}_{4,..,8}^{MM}$; the fifth choice incorporates just $\bar qq$ interpolators  ${\cal O}_{1,..,4}^{\bar qq}$ [see Eq. (\ref{swaveinterpolators})]. (b) $s$ wave, I=3/2: various choices of meson-meson interpolators ${\cal O}_{4,..,8}^{MM}$ [see Eq. (\ref{threehalf_swaveinterpolators})]. (c)  $p$ wave, I=1/2: levels with and without meson-meson interpolator ${\cal O}_6$ [see Eq. (\ref{pwaveinterpolators})]  in the basis.  }\label{fig:eff_eig_set_dep}
\end{figure*}

The dependence of the energy levels on the interpolator choice is summarized in Fig.~\ref{fig:eff_eig_set_dep}. We can clearly identify the relation between particular levels and the corresponding meson-meson interpolators, which will be detailed for each channel below.  

\begin{table*}[t]
\begin{ruledtabular}
\begin{tabular}{c|ccrclllllll}
& level  &interpol. & $t_0$& fit  & fit  &$aE_n=a\,\sqrt{s}$       &$E=\sqrt{s}$      & $\tfrac{\chi^2}{d.o.f.}$ 
 & $a\,p^*$  & $a^{2l}\, \rho_\ell^I(s)$&$\delta$\\
& $n$     &         &      & range & type & & [GeV] &  &  & & [degrees] \\
\hline
I=1/2 & 1 & ${\cal O}_{1,4,5,6,7,8}$ &4& 11-16&1 exp.  & 0.4768(28)  & 0.7593(45)  & 5.8/4 &$ \I\,$0.0889(31)  & 0.409(58)  & $ \I\,$28.2(5.8)\\
$s$ wave& 2 & ${\cal O}_{1,4,5,6,7,8}$ &2& 3-13 &2 exp.  & 0.777(13)   & 1.237(21)  & 9.6/7 &0.2835(87)         & 0.051(35)  &       82.0(5.3)\\
      & 3 & ${\cal O}_{1,2,5,6,8}$   &4& 6-11 &1 exp.  & 0.980(45)   & 1.561(72)  & 0.42/4  &0.410(26)         & -1.3(3.2) [*]&       162(28)\\
\hline
I=3/2 & 1  & ${\cal O}_{5,6,7,8}$   &4& 10-16&1 exp   & 0.5323(29)& 0.8478(46) & 0.018/5$~\dagger$& 0.0653(52)  &-1.67(22)  & -4.21(89)\\
$s$ wave& 2  & ${\cal O}_{5,6,8}$     &4& 8-13 &1 exp.     & 0.9979(74)& 1.589(12)  & 9.1/4   & 0.4208(43)  &-0.76(15)  &-29.1(4.9)\\
\hline
I=1/2 & 1  &${\cal O}_{1,2,3,6}$  &4& 8-16&1 exp.& 0.5749(19) & 0.9156(30) &10.7/7 &  0.1225(21)  & -0.0091(1)& 160.61(73)  \\    
$p$ wave& 2  &${\cal O}_{1,2,3,6}$  &4& 8-12&1 exp.& 0.9558(44) & 1.5223(70) &0.83/3 &  0.3958(26)  & -1.2(1.0)& 177.0(2.6)\\    
      &3  &${\cal O}_{1,2,3,5,6}$&4& 6-10&1 exp.& 1.080(11)  & 1.720(17)  &0.32/3 &  0.4686(65)  & -0.026(60)& 93.5(7.9)\\  
      & 4   &${\cal O}_{1,2,3,5,6}$&4& 6-10&1 exp.& 1.141(18)  & 1.817(28)  &1.2/3 &   0.503(10)   & 0.33(14) &  53(11)\\
\hline
I=3/2 & 1  & ${\cal O}_6$ & /  & 8-14&1 exp.& 0.9653(31) & 1.5356(48) & 4.5/5 & 0.4015(18)  &-0.443(91)& -8.6(1.8)    \\
$p$ wave&    &              &     &    &      &            &            &       &             &          & \\
\end{tabular}
\end{ruledtabular}
\caption{ 
Final results in four channels of $K\pi$ scattering: $s$ wave and $p$ wave with $I=1/2$ and $I=3/2$. The total momentum of the $K\pi$ system is $\mathbf{p}_\pi+\mathbf{p}_K=0$ in our simulation. For each $K\pi$ eigenstate, we present the energy $E=\sqrt{s}$, the momentum $p^*=|\mathbf{p}_\pi|=|\mathbf{p}_K|$ (\ref{pstar_q2}), the resulting scattering phase shift $\delta_\ell^I$ (\ref{zeta}) and the quantity $\rho_\ell^I$ defined in Eq. (\ref{def_rho}). The ``interpol'' column indicates which interpolators for the $s$ wave [see Eqs. (\ref{swaveinterpolators}, \ref{threehalf_swaveinterpolators})] and for the $p$ wave [see Eqs. (\ref{pwaveinterpolators}, \ref{threehalf_pwaveinterpolators})] are taken as our final choice in the variational basis. All fits are correlated with given $\chi^2$ (with exception of level $n=1$ for $I=3/2$ in $s$ wave, marked by $\dagger$).  
Note that the value of $\rho_\ell^I$ (\ref{def_rho}) has a huge error bar when $\delta$ is  $0^\circ$ or $180^\circ$ within error bar, as marked by [*].  Phase shifts are determined up to multiples of $180^\circ$ from Eq. (\ref{zeta}).
}\label{tab:phases}
\end{table*}

\begin{figure}[t]
\begin{center}
\includegraphics*[width=0.48\textwidth,clip]{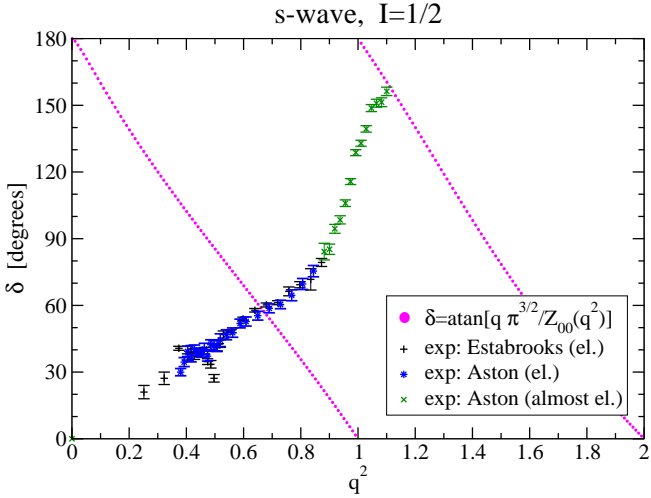}
\end{center}
\caption{
The analytic phase $\delta=\mathrm{atan}[\pi^{3/2}q/Z_{00}(q^2)]$ (see Eq. (\ref{zeta}))  as a function of $q^2$. The experimental phase $\delta_{\ell=0}^{I=1/2}$ \cite{Estabrooks:1977xe,Aston:1987ir} is also given as a function of $q^2=(p^*L/2\pi)^2$, where $L=16a$ is our lattice size and $p^*$ is the center-of-momentum frame of $\pi$ or $K$ in the  experiment. The energy levels   in a lattice simulation  arise for the values  of $q^2$ where the analytic and the ``experimental'' phases cross. Since the simulation is not performed at physical $m_{K,\pi}$ the crossings are slightly shifted in the actual simulation. 
   }\label{fig:luscher}
\end{figure}

\begin{figure*}[t]
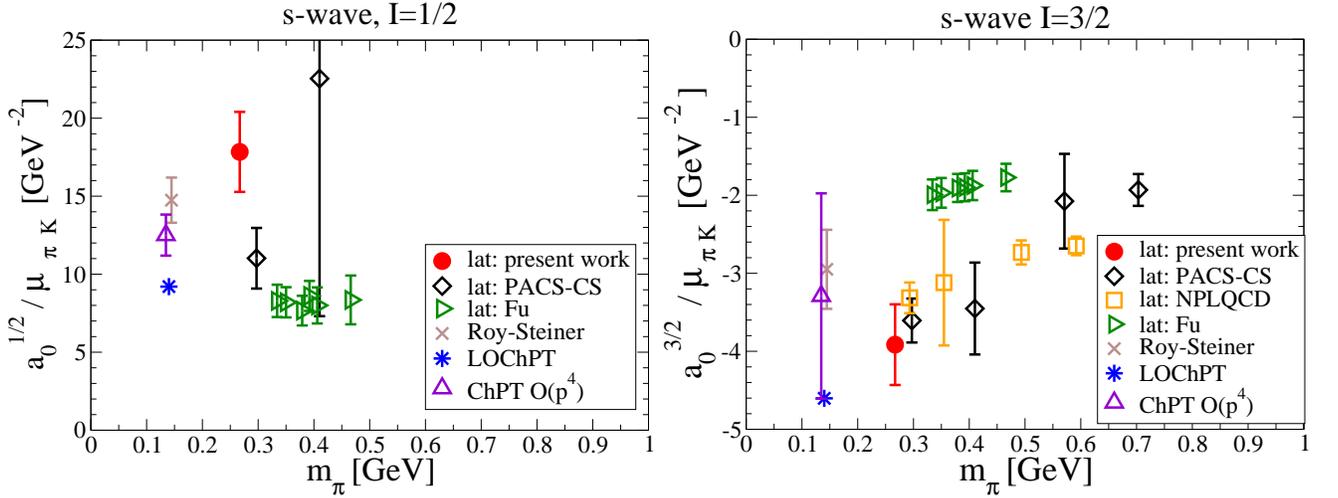

\begin{center}
\includegraphics*[width=0.48\textwidth,clip]{figs/isospin_1-2.eps}
\includegraphics*[width=0.48\textwidth,clip]{figs/isospin_3-2.eps}
\end{center}
\caption{The $s$ wave scattering lengths $a_0^I$ for $I=1/2$ and $3/2$ expressed as $a_0^I/\mu_{K\pi}$ which does not depend on $m_\pi$ at LOChPT. The result from the present simulation at a single $m_\pi\simeq 266~$MeV is compared with  results from other dynamical simulations \cite{Sasaki:2009cz,Beane:2006gj,Fu:2011wc} and with LOChPT, ChPT at ${\cal O}(p^4)$ \cite{Bernard:1990kw} and a Roy-Steiner approach \cite{Buettiker:2003pp}. 
   }\label{fig:a_scat}
\end{figure*}

\begin{figure}[t]
\begin{center}
\includegraphics*[width=0.48\textwidth,clip]{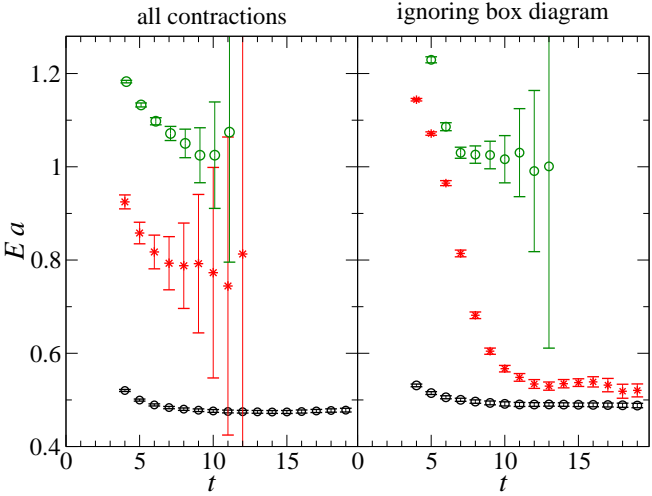}
\end{center}
\caption{Effective energies in the $s$ wave $I=1/2$ channel obtained from a $3\times 3$ correlation matrix using only meson-meson interpolators ${\cal O}_{5,7,8}$ (see (\ref{swaveinterpolators})) and $t_0=4$. Left: using all necessary contractions $C^{direct}+\tfrac{1}{2}C^{crossed}-\tfrac{3}{2}C^{box}$ listed in Appendix B and illustrated in Figs.~\ref{fig:contractions}a and  \ref{fig:contractions}c. Right: using only $C^{direct}+\tfrac{1}{2}C^{crossed}$ (Fig.~\ref{fig:contractions}a) and omitting the backtracking box contraction in Fig.~\ref{fig:contractions}c.  }\label{fig:ignore_dis}
\end{figure}

\subsubsection{$K\pi$ in $s$ wave, $I=1/2$}\label{sec:swave_half}

The experimental phase in Fig.~\ref{fig:phases} is positive and  rather slowly rising in this attractive channel. The scalar resonance $K^*_0(800)$ or $\kappa$ is controversial as it does not render  a typical Breit-Wigner shape with $\delta= 90^\circ$ at the position of the resonance. The experimental phase shift does not reach $\delta\simeq 90^\circ$ before $\sqrt{s}\simeq 1.3~$GeV which is in the vicinity of the $K_0^*(1430)$. 

In order to understand our lattice spectrum in this channel, we first compare the effective energies resulting from  
different subsets of interpolators ${\cal O}^{\bar qq}_{1,..,4}$ and 
 ${\cal O}^{MM}_{5,..,8}$ (\ref{swaveinterpolators}) which are plotted in Fig.~\ref{fig:eff_eig_set_dep}. 
The horizontal broken lines show the energies  of the 
noninteracting scattering states $K(n)\pi(-n)$. 
The dotted-dashed line corresponds to the noninteracting scattering state $K^*(0)\rho(0)$ with the energy $Ea=m_{\rho}a+m_{K^*}a\simeq 0.51+0.57=1.08$ ($m_{K^*}$ can be read off from  
Table \ref{tab:phases}, while $m_\rho$ is provided in Ref. \cite{Lang:2011mn}). The energy of the ground and the first excited states is  robust to the choice of the interpolator basis as long as meson-meson interpolators are included. Using just  $\bar q q$ interpolators ${\cal O}_{1,..,4}^{\bar qq}$ gives a much noisier ground state and higher energy for the first excited state (right  plot in Fig.~\ref{fig:eff_eig_set_dep}). If we omit ${\cal O}_6$ from the basis (\ref{swaveinterpolators}),  the state $K(1)\pi(-1)$  disappears from  the spectrum in Fig.~\ref{fig:eff_eig_set_dep} as expected. So the relatively noisy level $n=3$ (green circles at $Ea\simeq 0.95$) can be identified with the back-to-back momentum scattering state   $K(1)\pi(-1)$. 
If we omit ${\cal O}_7\simeq K^*(0)\rho(0)$, the corresponding state disappears from the spectrum, so level $n=4$  (blue left-facing triangles) can be identified with $K^*(0)\rho(0)$. 
Removal of ${\cal O}_7$ has little effect on the nearby levels (the resulting $E_{1,2,3}$ agree whether ${\cal O}_7$ is the basis or not): 
thus this interpolator appears to be weakly coupled to the system. 

Before presenting the values of the resulting phase shifts, we need to discuss up to which $E=\sqrt{s}$ the $K\pi$ scattering is elastic on our lattice, since relation (\ref{zeta}) rigorously applies only for elastic scattering. $K\pi$ in $s$ wave has $I(J^P)=\tfrac{1}{2}(0^+)$, which allows the low-lying scattering states $K^*(0)\rho(0)$, $K(0)\eta(0)$ and   $K(0)\eta'(0)$. Since our simulation has only dynamical $u/d$ quarks, only the heavier\footnote{The one that is not the Goldstone boson in  $SU(2)$, but is lifted by the chiral anomaly.}  $\eta_2$ can occur as an intermediate state in our simulation, and we are safe from inelasticity related to $K(0)\eta(0)$. The state $K(0)\eta_2(0)$ is expected to be heavier and we do not incorporate the corresponding interpolator, so we expect that our levels $n<4$ are not affected by that. Our level $n=4$ corresponds to the state $K^*(0)\rho(0)$, so the inelastic threshold definitely opens at $E_4 a \simeq 1.08$, while our levels $n<4$ are believed to be in the elastic regime. 

The effective energies and the final physics results for the three levels $n\leq 3$ in the elastic regime are presented in Fig.~\ref{fig:eff_eig_chosen} and Table \ref{tab:phases}, while the extracted phases are compared to experiment in Fig.~\ref{fig:phases}. The phase shift for the lowest level near threshold is imaginary and will be presented in terms of the scattering length below. The phase shifts in Fig.~\ref{fig:phases} arise from $n=2,3$ and agree  well with the experimental phase shifts. Note that both phase shifts are for $\sqrt{s}$ quite far away from threshold, so the exact position of the threshold $m_\pi+m_K$ in the lattice simulation is not expected to be of major importance here. 

The level $n=2$  corresponds to a phase shift value close to $90^\circ$, and we attribute this to the vicinity of the $K_0^*(1430)$ scalar resonance. Since $\delta$ for $n=3$ is already equal to $180^\circ$ within error bars, we cannot  provide a reliable estimate for the resonance mass and width of the $K_0^*(1430)$. Ignoring the huge error bar on $\rho_0^{1/2}(n=3)$, the central values of  $\rho_0^{1/2}(n=2,3)$ lead via a linear fit (\ref{BW_resonance}) to the resonance mass $m_R^{lat}\simeq 1.25~$ GeV and $\gamma^{lat}\simeq 0.67~$GeV$^2$. This rough estimate  of  $m_R^{lat}$ is within $\Gamma^{exp}=270(\pm 80)~$MeV from $m_R^{exp}=1425(\pm 50)~$MeV, while $\gamma^{lat}$ is also not far away from $\gamma^{exp}\simeq 0.83~$GeV$^2$ derived from $\Gamma^{exp}(K^*_{1430}\to K\pi)=(0.93 \pm 0.10)\cdot \Gamma^{exp}$. A reliable lattice determination of these two parameters in a future simulation will require an accurate value of $\delta$ near $K(1)\pi(-1)$ which is not $180^\circ$ within error bar.  
 
The  broad resonance $K^*(800)$ (or $\kappa$) does not lead to an additional energy state in our lattice spectrum. This is in agreement with expectation since the experimental phase shift never reaches $\delta\simeq 90^\circ$ near $\sqrt{s}\simeq m_\kappa$. Similar to the $\pi\pi$ $s$ wave various analyses assume, that there is a broad  
 $K^*(800)$ (or $\kappa$) resonance hidden behind the slowly rising phase shift below 1 GeV;
 it is associated with a resonance pole (in the second sheet) quite distant from the
 real axis. In order to understand that we cannot expect an additional level due to $\kappa$; we plot the analytic $\delta=\mathrm{atan}(q \pi^{3/2}/Z_{00})$ [see Eq. (\ref{zeta})] and the experimental phase $\delta_{0}^{1/2}$ as a function of $q^2=(p^* L/2\pi)^2$ with our $L=16a$ in Fig.~\ref{fig:luscher}. The energy states on the lattice are expected at those $q^2$ where analytic and experimental phases cross if $m_{\pi,K}$  is  physical. Since our $m_{\pi,K}$ are not physical, the crossings and energy levels are  slightly shifted in the actual simulation. The two crossings in Fig.~\ref{fig:luscher} correspond to the levels $n=2,\,3$ in Table  \ref{tab:phases}, while there is an additional crossing for $n=1$ below threshold at imaginary $\delta$ which is not plotted.  
Obviously, there is no crossing near   $\sqrt{s}\simeq m_\kappa$, since the lowest crossing in the plot\footnote{The crossing due to level $n=2$ takes place at $q^2=0.52(3)$ in the actual simulation. } at $q^2\simeq 0.7$ corresponds to $\sqrt{s}$ well above $1$ GeV.  
This indicates  that no energy level is to be expected near $\sqrt{s}\simeq m_\kappa$, just like confirmed by our simulation. We emphasize that the absence of an additional energy level near $\sqrt{s}\simeq m_\kappa$ in our simulation does not contradict the possible presence of the $\kappa$ pole in the second sheet; note that several analytical studies (for example Ref. \cite{Buettiker:2003pp,DescotesGenon:2006uk}) recover the experimental phase shift and do find the pole. 
The  lattice simulation is restricted to real $s$  and does not have direct access to look for poles as a function complex $s$ (just like experiment). So  our conclusion  is that we qualitatively agree with the experimental phase shift in this channel, but we cannot conclude whether the $\kappa$ pole exists or not.

Finally, we turn our attention to the lowest level slightly below threshold, which has small and imaginary $p^*$ and $\delta$. It allows the extraction of the scattering length  from $\rho_0^{1/2}(n=1)$ in Table \ref{tab:phases}  according to \eq{effective_range} \footnote{Only the first term on the right-hand side of Eq. \eq{effective_range} is used to determine $a_{0}^{1/2}$ as the second term is much smaller due to small $p^{*}$, and can safely be neglected.}
\begin{align}
a^{1/2}_0&=5.13\pm 0.73\, a=0.636\pm 0.090~\mathrm{fm} \\
 \frac{a^{1/2}_0}{\mu_{K\pi}}&=17.9\pm 2.5~\mathrm{GeV}^{-2} \quad \mathrm{at}\quad  m_\pi\simeq 266~ \mathrm{MeV}\nonumber
\end{align}
where $m_{K,\pi}$ from  the simulation were inserted to the reduced mass $\mu_{K\pi}$. In Fig.~\ref{fig:a_scat} we compare the values of the ratio $a^{I=1/2}_0/\mu_{K\pi}$ with other dynamical lattice simulations. We choose to present this ratio as it is independent of $m_{K,\pi}$ in LOChPT
\begin{equation}
\frac{a^{1/2}_0}{\mu_{k\pi}}=-2 ~\frac{a^{3/2}_0}{\mu_{k\pi}} =\frac{1}{2\pi F_\pi^2}~[1+O(m_\pi^2)]~\FC
\end{equation}
with  $F_\pi\simeq 0.13~$GeV. 
This leading-order prediction is also shown in Fig.~\ref{fig:a_scat} together with the result from ChPT at $O(p^4)$ \cite{Bernard:1990kw} and  the result using a Roy-Steiner analysis \cite{Buettiker:2003pp}. We are not able to perform the extrapolation of our $a_0^{I=1/2}$ to physical pion mass as we calculated its  value at only one $m_\pi$.  

\vspace{0.2cm}

As indicated above, we find only one energy level below $E=1~$GeV in our present simulation: this level is related to $K(0)\pi(0)$, while $\kappa$ does not lead to an additional energy level. In our previous simulation \cite{Prelovsek:2009bk,Prelovsek:2010kg}, we employed only four-quark interpolators $\sum_{\mathbf{x}}\bar q(\mathbf{x})q(\mathbf{x})\bar q(\mathbf{x})q(\mathbf{x})$ with 5 different color and Dirac structures, which all had $I=1/2$ and $J^P=0^+$. The necessary contractions are shown in Fig.~\ref{fig:contractions}a and  \ref{fig:contractions}c. We omitted\footnote{The first reason for this omission was the numerical cost. The second reason had physical motivation of artificially prohibiting the mixing $\bar qq\bar qq\to \bar qq\to \bar qq\bar qq$, so that a $\bar qq\bar qq$ Fock component could be attributed to the resulting state.} the backtracking box contraction \ref{fig:contractions}c in Refs. \cite{Prelovsek:2009bk,Prelovsek:2010kg} and the simulation rendered an additional state near $K(0)\pi(0)$ which was attributed to $\kappa$ with a sizable tetraquark Fock component in Refs. \cite{Prelovsek:2009bk,Prelovsek:2010kg}. The effect of the box contraction \ref{fig:contractions}c in our present simulation is shown in Fig.~\ref{fig:ignore_dis}, where the spectrum is calculated using only $(\bar qq)(\bar qq)$ interpolators ${\cal O}_{5,7,8}$ [see Eq. (\ref{swaveinterpolators})]; these are similar to  ${\cal O}_{1,2,3}$ used in Refs. \cite{Prelovsek:2009bk,Prelovsek:2010kg}. There is only one energy state below $E=1~$GeV when all necessary contractions (\ref{fig:contractions}a and \ref{fig:contractions}c) are incorporated, which agrees with the result in Fig.~\ref{fig:eff_eig_chosen} and with our conclusions above.  However, if the backtracking box contraction \ref{fig:contractions}c is neglected, an additional energy level near $K(0)\pi(0)$ appears. A proper quantum field theory treatment requires incorporation of all Wick contractions, so the additional level seems to be an artifact of the  approximation used in \cite{Prelovsek:2009bk,Prelovsek:2010kg}. This interesting observation may be fruitful in trying to understand the physics of light scalar mesons in  future explorations.

\subsubsection{$K\pi$ in $s$ wave, $I=3/2$}
This $K^+\pi^+$ channel is repulsive, and the experimental phase shift  in Fig.~\ref{fig:phases} is negative and slowly rising. A negative phase shift in Table \ref{tab:phases} is also 
 observed for the lowest two states in our lattice simulation, since they are clearly above noninteracting $K(0)\pi(0)$ and $K(1)\pi(-1)$ in Fig.~\ref{fig:eff_eig_chosen}. The value of $\delta \simeq -30^\circ$  at $\sqrt{s}\simeq 1.6~$GeV agrees nicely  the experiment.

We do not see any additional state between these two levels, which agrees with the fact that resonances have  not been experimentally observed in this repulsive channel.

Investigation of the spectrum with different interpolator choices in Fig.~\ref{fig:eff_eig_set_dep} indicates that ${\cal O}_6$ [see Eq. (\ref{threehalf_swaveinterpolators})] is responsible for level $n=2$, and ${\cal O}_7\sim K^*(0)\rho(0)$ is responsible for the level $n=3$. The $K\eta$ and $K\eta'$ do not contribute to the $I=3/2$ channel due to isospin, so the lattice data and the experimental data are completely elastic up to rather high $\sqrt{s}=m_{K^*}+m_\rho$.  

The lowest level slightly above threshold allows the extraction of the scattering length  from $\rho_0^{3/2}(n=1)$ in Table \ref{tab:phases}  according to \eq{effective_range} 
\begin{align}
a^{I=3/2}_0&=-1.13\pm 0.15\, a=-0.140\pm 0.018~\mathrm{fm} \\
 \frac{a^{I=3/2}_0}{\mu_{K\pi}}&=-3.94\pm 0.52~\mathrm{GeV}^{-2} \quad \mathrm{at}\quad  m_\pi\simeq 266~ \mathrm{MeV}~.\nonumber
\end{align}
This scattering length is compared with other lattice and continuum results in Fig.~\ref{fig:a_scat}. 

Due to the smooth behavior of the phase shift observed in experiment \cite{Estabrooks:1977xe}, we attempt to estimate also the  effective  range $r_0^{3/2}$  by employing the effective range formula (\ref{effective_range}). From the values of $\rho_0^{3/2}$, $p^*$ and $\sqrt{s}$ for the levels $n=1,2$ in Table \ref{tab:phases}, we extract 

\begin{align}
a^{3/2}_0 &=-1.12\pm 0.15\,a  = -0.139\pm 0.018~ \mathrm{fm} \\
r_0^{I=3/2} &=1.5\pm 2.0 \,a = 0.19\pm 0.25~ \mathrm{fm} \nonumber
\end{align}
at our $m_\pi$, which indicates that the dependence of $p^* \cot \delta$ (\ref{effective_range}) on $p^*$ is small (zero within errors) up to $p^*\simeq 0.67~$GeV. Our $I=3/2$ phase shift is therefore dictated by the scattering length for $p^*$ as high as  $p^*\simeq 0.67~$GeV. 
  The experimental effective range at physical pion mass is  
$r^{3/2}_0=-0.346(\pm 0.060)~$fm \cite{Estabrooks:1977xe}.

\subsubsection{$K\pi$ in $p$ wave, $I=1/2$}

The elastic region is dominated by a vanilla-style resonance: the $K^*(892)$
with  a width of $\approx 50~$MeV from experiments. Experiments indicate further resonances   $K^*(1410)$ and $K^*(1680)$, where the first one is not established in all experiments. 

The spectrum on our lattice is shown in Fig.~\ref{fig:eff_eig_chosen}, and all these levels are expected to be in the elastic regime, as discussed below. 
The scattering level $K(1)\pi(-1)$ is seen only if the meson-meson interpolator $O_6$ [see Eq. (\ref{pwaveinterpolators})] is taken in the interpolator basis, while $\bar qq$ interpolators alone do not render it (see Fig.~\ref{fig:eff_eig_set_dep}). The other three levels (red stars and pink crosses) appear  away from the noninteracting scattering states $K(n)\pi(-n)$ and are  candidates to be related with the resonances $K^*(892),~ K^*(1410),~ K^*(1680)$. 

The ground state is due to the $K^*(892)$ and gives a rather high $\delta\simeq 160^\circ$ at $\sqrt{s}\simeq 915~ \mathrm{MeV}\simeq m_{K^*}$ (see Fig.~\ref{fig:phases}). This is not surprising since the phase should be  rising very steeply 
as expected from the narrow width  $\Gamma^{lat}\simeq (p_{lat}^*/p_{exp}^*)^3\Gamma^{exp}\simeq (0.19/0.29)^3 \cdot 50~\mathrm{MeV}\simeq 14~$MeV derived from $\Gamma^{exp}$ and assuming the same coupling $\gamma$ [see Eq. (\ref{Gamma})] in both cases. Since we have only one value of the  phase shift near $\sqrt{s}\simeq m_{K^*}$, we cannot rigorously extract the $K^*(892)$ mass or width, but we expect that the phase shift would pass $90^\circ$ at about $m_R^{lat}=(s+\sqrt{s}\Gamma^{lat}\cot\delta)^{1/2}\simeq 896~$MeV (\ref{BW_resonance}), where the derived  width $\Gamma^{lat}$  quoted above was assumed. 
Extraction of the $K^*(892)$ width in future lattice simulations will  be possible only if two phase shifts are extracted in close vicinity of $\sqrt{s}\simeq m_{K^*}$.  Simulations at nonzero-total momentum and the relevant phase shift formulae for the $p$ wave \cite{Leskovec:2012gb,Doring:2012eu,Gockeler:2012yj} might come to the rescue in this case. 

The third and fourth level at $E_3=1.720(17)~$GeV  and $E_4=1.817(28)~$GeV  are most probably related to the wide resonances $K^*(1410)~, K^*(1680)$ which are the only $p$ wave resonances which appear  between $1$ and $2$ GeV in experiment.  However, the level  $E_3$ appears too high in comparison to the $K^*(1410)$ 
even if one takes into consideration its large experimental width $\Gamma^{exp}=232(21)~$MeV and our unphysical $m_{\pi,K}$. Notice, however, that in a small box and with unphysically heavy u/d quarks the situation is quite different from experiment, where both the $K^*(1410)$ and the $K^*(1680)$ have a sizable branching ratio into the $K^*(892)\pi$ and $\rho K$ channels. The inelastic threshold in this $J^P=1^-$ channel opens at $K^*(1)\pi(-1)$ or $K(1)\rho(-1)$ in $p$ wave. This is at rather   high $E\simeq 1.9$ GeV, so all levels (except possibly $n=4$) are  in the elastic regime in our simulation. Therefore, our situation is somewhat unphysical in this case, as we can only consider elastic scattering in the $K \pi$ channel. Our results are consistent with the observations in the simulation \cite{Engel:2011aa}, where the energy level associated with the $K^*(1410)$ is also observed substantially higher than the physical state. Notice that for our kinematics, one expects a phase shift which is monotonically increasing in the vicinity of the levels $E_3$ and $E_4$. This is not in conflict with Fig.~\ref{fig:phases}, where the phase is restricted to $0<\delta<180^\circ$, since the phase obtained from Eq. (\ref{zeta}) is undetermined up to multiples of $180^\circ$.

\subsubsection{$K\pi$ in $p$ wave, $I=3/2$}

The experimental phase in this repulsive channel is negative and very small. 

On the lattice, we extract the phase from a single level given by the  correlator $\langle {\cal O}_6(t)|{\cal O}_6^\dagger(0)\rangle \to \E^{-E t}$ [see Eq. (\ref{threehalf_pwaveinterpolators})]. 
 It appears slightly above the  noninteracting $K(1)\pi(-1)$ and  renders a negative phase $\delta^{lat}(\sqrt{s}\simeq 1.5~GeV)=-8.6^\circ\pm 1.8^\circ$. This is in qualitative agreement with the experimental phase which is negative and   
 does not exceed $-10^\circ$ for $\sqrt{s}$ as large as $1.8~$GeV  (see Fig.~\ref{fig:phases}), although both phases do not agree within errors.

\section{Summary and outlook}
We simulated $K\pi$ scattering with lattice QCD and  extracted the  elastic phase shifts  $\delta_\ell^I$  in $s$ wave and $p$ wave for $I=1/2,3/2$ at several values of the $K\pi$ invariant mass $\sqrt{s}$.  We used a single lattice QCD ensemble of size $L\simeq 2~$fm with dynamical $u$ and $d$ quarks. Our results for phase shifts and scattering lengths apply for the values of $m_\pi \simeq 266~$MeV and $m_K\simeq 552~$MeV. The total three-momentum of the $K\pi$ system is zero in our simulation.

First, we extracted the energy levels of the $K\pi$ system on our finite lattice. In all channels, we observe the expected $K(n)\pi(-n)$ scattering state levels, which are shifted relative to the noninteractive case due to the interaction. In both attractive $I=1/2$ channels, we observe additional levels which are related to resonances   $K_0^*(1430)$ in $s$ wave and $K^*(892)$, $K^*(1400)$ and $K^*(1680)$ in $p$ wave. 

The phase shifts are extracted from the energy levels using L\" uscher's method. They are compared to the experimental phase shifts  in Fig.~\ref{fig:phases} and exhibit 
qualitative agreement in all four channels:  
\begin{itemize}
\item
\underline{ $s$ wave, $I\!=\!1/2$}: The phase is positive and 
yields the scattering length $a_0^{I=1/2}=0.636(90)~$fm at our $m_\pi$. 
Our first excited state is observed with $\delta \simeq 90^\circ$ near $K_0^*(1430)$, 
which implies that the phase does not reach $90^\circ$ below $\sqrt{s}=1~$GeV in our simulation.  
This agrees with the experimental finding that the phase is {\it not } $90^\circ $ near $\sqrt{s}\simeq m_\kappa$ and that the controversial $\kappa$ resonance cannot be described by a conventional Breit-Wigner shape. 
\item
\underline{ $s$ wave, $I\!=\!3/2$}: The phase is negative and reaches $\delta_{0}^{I=3/2}\simeq -30^\circ$ at $\sqrt{s}\simeq 1.6~$GeV in agreement with experiment. We extract  
$a_0^{I=3/2}=-0.140(18)~$fm and we find very mild dependence of $\sqrt{s}\cot \delta$ on $p^*$ up to $p^*=0.67~$GeV. 
\item
\underline{ $p$ wave, $I\!=\!1/2$}: Our spectrum and the phases favor the existence of three resonances  $K^*(892)$, $K^*(1410)$ and $K^*(1680)$ below $2~$GeV, but our energy level for  $K^*(1410)$ is higher than the experimental one. We did not extract  the $K^*(892)$  width as we have only one energy level in the vicinity of this well-established narrow resonance. 
\item
\underline{ $p$ wave, $I\!=\!3/2$}: The phase is negative and very small, which is observed also in  experiment. 
\end{itemize}

The extraction of the widths for  the resonances in the $I=1/2$ channels is left for future simulation. This will require at least two   values of the phases $\delta(\sqrt{s})$ in the vicinity $\sqrt{s}\simeq m_R\pm \Gamma_R$ of each resonance, which is particularly challenging for a narrow resonance like the $K^*(892)$. This might be possible to achieve for the $p$ wave using simulations 
with $\mathbf{P}=\mathbf{p}_\pi+\mathbf{p}_K\not =0$  which would provide additional values of phases  $\delta_{\ell=1}$ at $s=E^2-\mathbf{P}^2$.  The relations which allow an extraction of $\delta_{\ell=1}$ from the energies  in this case are derived in Refs. \cite{Leskovec:2012gb,Doring:2012eu,Gockeler:2012yj}, while sample interpolators are explicitly listed in Ref. \cite{Leskovec:2012gb}. This will be much more challenging for the $s$ wave since $\delta_{\ell=0}$ is always mixed with $\delta_{\ell=1}$ in the phase shift relations 
 for the $A_1$ irreducible representation when $\mathbf{P}\not =0$ and $m_\pi\not = m_K$  \cite{Fu:2011xz,Leskovec:2012gb,Doring:2012eu,Gockeler:2012yj}. 
The extraction of the $s$-wave phase shifts is therefore more reliable with the present simulation 
 at  $\mathbf{P}=0$ and calls for similar simulations at different lattice sizes.

\acknowledgments
First of all, we would like to thank Anna Hasenfratz for providing the
gauge configurations used for this work.  We would like to thank S. Descotes-Genon, M.~D\"oring, A.~Rusetsky and R.~M.~Woloshyn for valuable discussions. The calculations have been performed on the theory
cluster at TRIUMF and on local clusters at the Universities of Graz and
Ljubljana. We thank these institutions for providing the support. This work is
supported by the Slovenian Research Agency and by the Natural Sciences and Engineering Research Council of Canada (NSERC).

\newpage
\begin{appendix}
\section{Interpolators}\label{app_a}
\subsection{Single pion and kaon interpolators}

For the single pion or single kaon sectors we have $6$ interpolators, using three
smearing widths [see Eq. (\ref{smearing})] for each of the two Dirac structures,
\begin{eqnarray} \label{pion_and_kaon_interpolators} 
{\cal O}^\pi_{type,s}(\mathbf{p},t)
&=&\sum _{\mathbf{x}}\dbar_s(x) \Gamma_{type}\E^{\I\mathbf{p}\mathbf{x}} u_s(x)\FC\nonumber\\
{\cal O}^K_{type,s}(\mathbf{p},t)
&=&\sum _{\mathbf{x}}\sbar_s(x) \Gamma_{type}\E^{\I\mathbf{p}\mathbf{x}} u_s(x)\FC\\
\Gamma_1&=&\gamma_5,\quad \Gamma_2=\gamma_5\gamma_t,\quad s=n,\,m,\,w \nonumber\FD
\end{eqnarray}  
Here we determine the energy levels for different values of the total  momentum
$\mathbf{p}$ in order to study the dispersion relation.

\subsection{Interpolators for the $K\pi$ system with $\mathbf{P}=0$}

For the $K\pi$ system we  use $\bar qq$ and meson-meson ($K\pi$, $\rho K^*$, $K_1a_1$) interpolators with appropriate quantum numbers. The interpolators for all four channels are listed in the following subsections. 

The meson-meson interpolators are expressed as products of two meson currents, where each  meson current $M(\mathbf{p})$ has momentum $\mathbf{p}$ projected to $0$ or  $\tfrac{2\pi}{L}\mathbf{e}_i$   
\begin{align}
\label{mom}
M(\mathbf{p})&\equiv \sum_{\mathbf{x}} \bar q_1(\mathbf{x},t) \Gamma \E^{i \mathbf{p}\mathbf{x}} q_2(\mathbf{x},t)\nonumber\\
\mathbf{p}=&
\begin{cases}
\mathbf{0} &   \\
 \mathbf{p}_i\equiv \tfrac{2\pi}{L} \mathbf{e}_i & \quad i=x,y,z ~.
\end{cases}
\end{align}
We use the following  flavor combinations and $\Gamma$ matrices for $M(\mathbf{p})$
\begin{align}
\pi^+&=\dbar \gamma_5 u \ , \  \ \quad \pi^0=\sqff{1}{2}\, (\ubar  \gamma_5 u-\dbar  \gamma_5 d)\nonumber\\
K^+ &=\sbar  \gamma_5 u \ ,\  \ \quad K^0 =\sbar \gamma_5 d\nonumber\\
(\rho^+)_i&=\dbar \gamma_i u \ ,\ \  \quad (\rho^0)_i=\sqff{1}{2}\, (\ubar  \gamma_i u-\dbar  \gamma_i d)\nonumber\\
(K^{*+})_i &=\sbar  \gamma_i u \ ,\ \  \quad (K^{*0})_i =\sbar \gamma_i d\nonumber\\
(a_1^+)_i&=\dbar \gamma_i \gamma_5 u \ , \quad (a_1^0)_i=\sqff{1}{2}\, (\ubar  \gamma_i \gamma_5 u-\dbar  \gamma_i \gamma_5 d)\nonumber\\
(K_1^{+})_i &=\sbar  \gamma_i \gamma_5u \ , \quad (K_1^{0})_i =\sbar \gamma_i \gamma_5 d\nonumber
\end{align}
where $i=x,y,z$ refers to the three spatial directions. 
Each quark $q_s$ is smeared according to Eq. (\ref{smearing}) with the number of eigenvectors $N_v$, provided for the interpolators below.   

We combine these meson currents into the $I=1/2$ combination
\begin{eqnarray}
\left\vert I ,\,I_3 \right\rangle=\left\vert \ot ,\,\ot\right\rangle&=& \sqff{1}{3}\, K^+\, \pi^0+\sqff{2}{3}\, K^0\, \pi^+  
\end{eqnarray}
or the $I=3/2$ combination
 \begin{equation}
\left\vert I ,\,I_3 \right\rangle=\left\vert \ff{3}{2} ,\,\ff{3}{2}\right\rangle = K^+\,\pi^+
\end{equation}
and analogously for $(\rho,K^*)$ and $(a_1,K_1)$ pairs which carry  different $J^P$ and the same isospin. 
 
Our quark-antiquark interpolators  below contain also a covariant derivative, defined as 
\begin{equation}
\label{cov_deriv}
\overrightarrow{\nabla}_i(\mathbf{x},\mathbf{y})=U_i(\mathbf{x},0)
\delta_{\mathbf{x}+\mathbf{i}, \mathbf{y}}-U_i^\dagger(\mathbf{x}-\mathbf{i},0)
\delta_{\mathbf{x}-\mathbf{i}, \mathbf{y}}\FD
\end{equation} 
 It acts on the spatial
and color indices and leaves time and Dirac indices intact. 


\subsection{$K\pi$ in $s$ wave, $I=1/2$}

For the  $\kappa$ channel we employ $4$ quark-antiquark interpolators and $4$ meson-meson
interpolators in the variational basis. Interpolators ${\cal O}_{1-5}$ are built using $q_m$ with $N_v=64$ eigenvectors, while interpolators ${\cal O}_{6-8}$ take $q_w$ with $32$ eigenvectors due to the 
sizable numerical cost related to them.  
The quark-antiquark
interpolators  ${\cal O}_{1-4}$ 
differ in Dirac and color structure. The interpolator ${\cal O}_{5}$ is a $ K
\pi$ interpolator where both pseudoscalars are at rest, whereas for ${\cal
O}_{6}$ they have oppositely oriented unit momentum, summed over all spatial
directions. Finally ${\cal O}_{6}$ and ${\cal O}_{7}$ are $\rho K^*$ and $a_1
K_1$ at rest. These two are in the inelastic region and their (ir)relevance is
discussed in Sec. \ref{sec:swave_half}. So we compute an $8\times 8$ correlation matrix with 
\begin{align}
\label{swaveinterpolators}
{\cal O}_{1}&=\sum _{\mathbf{x}}\,\sbar(x)\,u(x)\FC\\
{\cal O}_{2}&=\sum _{\mathbf{x},i}\,\sbar(x)\,\gamma_i \overrightarrow{\nabla}_i\,u(x)\FC\nonumber\\
{\cal O}_{3}&=\sum _{\mathbf{x},i}\,\sbar(x)\,\gamma_t\,\gamma_i \overrightarrow{\nabla}_i\,u(x)\FC\nonumber\\
{\cal O}_{4}&=\sum_{\mathbf{x},i}\,\sbar(x)\,
\overleftarrow{\nabla}_i\,\overrightarrow{\nabla}_i\,u(x)\FC\nonumber\\
{\cal O}_{5}&=
\sqff{1}{3} K ^+(\mathbf{0})\pi^0(\mathbf{0})+\sqff{2}{3} K ^0(\mathbf{0})\pi^+(\mathbf{0})\FC\nonumber\\
{\cal O}_{6}&=\sum_{i} 
\left[\sqff{1}{3} K ^+(\mathbf{p}_i)\pi^0(-\mathbf{p}_i)
+\sqff{2}{3} K^0(\mathbf{p}_i)\pi^+(-\mathbf{p}_i)\right]\nonumber\\
&\qquad\qquad+(\mathbf{p}_i\leftrightarrow -\mathbf{p}_i)\FC\nonumber\\
{\cal O}_{7}&=\sum_{i}  \left[\sqff{1}{3} K ^{*+}_i(\mathbf{0})\rho^0_i(\mathbf{0})
+\sqff{2}{3} K ^{*0}_i(\mathbf{0})\rho^+_i(\mathbf{0})\right]\FC\nonumber\\
{\cal O}_{8}&=\sum_{i}  
\left[\sqff{1}{3}\mathrm{K_1}^{+}_i(\mathbf{0})\, {a_1^0}_i(\mathbf{0})+\sqff{2}{3}\mathrm{K_1}^{0}_i(\mathbf{0})\, {a_1^+}_i(\mathbf{0})\right]\FC\nonumber
\end{align}
and the sum on $i$ runs over $i=x,y,z$.  The momenta $\mathbf{p}_i$ are given in Eq. (\ref{mom}).

\subsection{$K\pi$ in $s$ wave, $I=3/2$}

For the  exotic $I=\frac{3}{2}$ channel we use the corresponding interpolators ${\cal O}_{5-8}$ (\ref{swaveinterpolators}) and the same choice of smearings, just a different isospin projection 
\begin{align}\label{threehalf_swaveinterpolators}
{\cal O}_{5}&=
 K ^+(\mathbf{0})\pi^+(\mathbf{0})\FC \\
{\cal O}_{6}&=\sum_{i} K ^+(\mathbf{p}_i)\pi^+(-\mathbf{p}_i) + K ^+(-\mathbf{p}_i)\pi^+(\mathbf{p}_i)\FC\nonumber\\
{\cal O}_{7}&=\sum_{i}  K ^{*+}_i(\mathbf{0})\rho^+_i(\mathbf{0})\FC\nonumber\\
{\cal O}_{8}&=\sum_{i}  \mathrm{K_1}^{+}_i(\mathbf{0})\, {a_1^+}_i(\mathbf{0})\FD\nonumber
\end{align}
The naming scheme is kept analogous to Eq. (\ref{swaveinterpolators}).


\subsection{$K\pi$ in $p$ wave, $I=1/2$}

For the $K^*$ channel we
employ quark-antiquark interpolators ${\cal O}_{1-5}$ and 
one kaon-pion interpolator  ${\cal O}_{6}$, which are all built using $q_n$ with $N_v=96$ 
\begin{align}
\label{pwaveinterpolators}
{\cal O}_{1,i}&=\sum _{\mathbf{x}}\,\sbar(x)\,\gamma_i \,u(x)\FC\\
{\cal O}_{2,i}&=\sum _{\mathbf{x}}\,\sbar(x)\,\gamma_t\gamma_i \,u(x)\FC\nonumber\\
{\cal O}_{3,i}&=\sum_{\mathbf{x},j}\,\sbar(x)\,
\overleftarrow{\nabla}_j\,\gamma_i\,\overrightarrow{\nabla}_j\,u(x)\FC\nonumber\\
{\cal O}_{4,i}&=\sum_{\mathbf{x}}\,\sbar(x)\,
\tfrac{1}{2} \left[\overrightarrow{\nabla}_i -\overleftarrow{\nabla}_i\right]\,u(x) 
\FC\nonumber\\
{\cal O}_{5,i}&=\sum_{\mathbf{x},j,k} \,\epsilon_{ijk} 
 \,\sbar(x)\gamma_j\gamma_5 \,\tfrac{1}{2}
\left[\overrightarrow{\nabla}_k  -\overleftarrow{\nabla}_k\right] 
\, u(x)\FC\nonumber\\
{\cal O}_{6,i}&=
\sqff{1}{3} K^+(\mathbf{p}_i)\pi^0(-\mathbf{p}_i)
+\sqff{2}{3} K^0(\mathbf{p}_i)\pi^+(-\mathbf{p}_i)\nonumber\\
&\qquad -(\mathbf{p}_i\leftrightarrow -\mathbf{p}_i)\FD\nonumber
\end{align}
Here, the open polarization index is $i=x,y,z$, and we average the resulting correlation matrices over three polarizations. 
The linear combinations of the derivatives
in ${\cal O}_{4,5}$ render  good $C$ parity in the $SU(3)$ flavor limit.
Although $C$ is not a good quantum
number due to $m_s\not = m_{u,d}$, such a combination is advantageous as discussed, e.g., in Ref.
\cite{Engel:2011aa}.

\subsection{$K\pi$ in $p$ wave, $I=3/2$}

For the  $I=3/2$ $p$ wave channel, we use only ${\cal O}_6$ 
from Eq. (\ref{pwaveinterpolators}) with appropriate choice of flavor
\begin{equation}\label{threehalf_pwaveinterpolators}
{\cal O}_{6,i}=
K ^+(\mathbf{p}_i)\pi^+(-\mathbf{p}_i)- K ^+(-\mathbf{p}_i)\pi^+(\mathbf{p}_i)~,
\end{equation}
where $i=x,y,z$ is the polarization index. 
As there is just one interpolator in this case, we present its complete expression for  convenience
\begin{align*}
{\cal O}_{6,i}&=\sum_{\mathbf{x}_1,\mathbf{x}_2} [\E^{\I (\mathbf{p}_i \mathbf{x}_1-\mathbf{p}_i \mathbf{x}_2)}\, 
\bar s_n(t,\mathbf{x}_1)\gamma_{5} u_n(t,\mathbf{x}_1)~ \nonumber\\
&\times \bar d_n(t,\mathbf{x}_2)\gamma_{5} u_n(t,\mathbf{x}_2)]\quad  -(\mathbf{p}_i\leftrightarrow -\mathbf{p}_i)\FD
\end{align*}
Other employed meson-meson interpolators can be expressed in terms of the  quark fields in an analogous way. 

\section{Wick contractions}\label{app_b}

\begin{figure}[t]
\begin{center}
\includegraphics*[width=0.4\textwidth,clip]{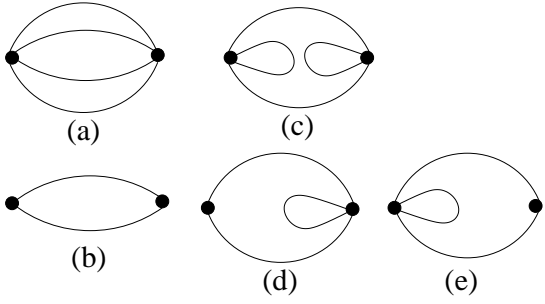}
\end{center}
\caption{
Contractions for our correlators with $\bar qq$ and meson-meson interpolators. 
Only (a) appears for $I=3/2$, while all these contractions appear for $I=1/2$. 
The ``backtracking'' contractions (c) and (d) require an all-to-all method.}\label{fig:contractions}
\end{figure}

Here, we provide the expressions for the $I=1/2$ and $I=3/2$ contractions. In the distillation method, they are expressed in terms of the perambulators for the light quark $\tau_u$ and the strange quark $\tau_s$,   as well as $\phi$  matrices which depend on the shape functions ${\cal F}$. We use exactly the same definitions of these quantities as in Sec. IIC and Appendix A of Ref. \cite{Lang:2011mn}, so we omit the definitions here.   

Any  annihilation operator  ${\cal O}(t)$ above   can be expressed in terms of 
\begin{align}
&{\cal O}^{MM}(t) = \sqrt{\tfrac{1}{3}}~\bar{s}(t) \Gamma_2 {\cal F}(p_2)u(t) \\
&\times \sqrt{\tfrac{1}{2}}~[\bar{u}(t) \Gamma_{1} {\cal F}(p_1)u(t)- \bar{d}(t) \Gamma_{1} {\cal F}(p_1)d(t) ]\nonumber\\
&+\sqrt{\tfrac{2}{3}}~ \bar{s}(t) \Gamma_2{\cal F}(p_2) d(t) ~ \bar{d}(t) \Gamma_{1}{\cal F}(p_1) u(t)\FC\nonumber\\
&{\cal O}_{\bar qq}(t) = \bar{s}(t)~ \Gamma_0 ~{\cal F}_0(P)~ u(t)~, \nonumber
\end{align}
and  creation operators ${\cal O}(t')$  at source can be expressed as 
\begin{align}
&{\cal O}_{MM}^{\dagger}(t') = C_{MM}  \bigl[\sqrt{\tfrac{1}{3}}~ \bar{u}(t') \Gamma_2' {\cal F}(-p_2') s(t')\\
&\times \sqrt{\tfrac{1}{2}}~[\bar{u}(t') \Gamma_{1}' {\cal F}(-p_1')u(t') - \bar{d}(t') \Gamma_{1}' {\cal F}(-p_1')d(t')]\nonumber\\
&+\sqrt{\tfrac{2}{3}}~\bar{d}(t') \Gamma_2' {\cal F}(-p_2')s(t')~\bar{u}(t') \Gamma_{1}' {\cal F}(-p_1') d(t')\bigr]\FC\nonumber\\
&{\cal O}_{\bar qq}^{\dagger}(t') = C_{\bar qq}~\bar{u}(t') ~\Gamma_0' ~{\cal F}_0'(-P)~s(t')\FD\nonumber
\end{align}

The $I=3/2$ case involves only meson-meson interpolators and only the connected contractions in Fig.~\ref{fig:contractions}a
\begin{equation}
\langle {\cal O}_{3/2}^{MM}(t)|{\cal O}_{3/2 }^{MM\dagger}(t')\rangle =  C_{MM} ~\bigl[ C^{direct}(t,t')-C^{crossed}(t,t')\bigr]\FC
\end{equation}
where separate terms are explicitly given below. 

The $I=1/2$ case involves all contributions depicted in Fig.~\ref{fig:contractions}
\begin{align}
&\langle {\cal O}_{1/2}^{MM}(t)|{\cal O}_{1/2 }^{MM\dagger}(t')\rangle =  C_{MM} ~\bigl[ C^{direct}(t,t')\\
&+\tfrac{1}{2}~C^{crossed}(t,t')-\tfrac{3}{2}~C^{box}(t,t')\bigr]\FC\nonumber\\
&\langle {\cal O}_{\bar qq}(t)|{\cal O}_{\bar qq}^\dagger(t')\rangle  = -C_{\bar qq} \nonumber\\
&\times   Tr[\tau_{s}(t',t) \Gamma_0 \phi(t,{\cal F}_0(p)) ~\tau_{u}(t,t') \Gamma_0'  \phi(t',{\cal F}_0'(-p)]\FC\nonumber\\
&\langle {\cal O}_{1/2}^{MM}(t)|{\cal O}^{\bar qq\dagger}(t')\rangle = C_{MM}~\sqrt{\tfrac{3}{2}} ~Tr[\tau_{s}(t',t) \Gamma_2 \phi(t,{\cal F}(p_2))\nonumber\\
&\times  ~\tau_{u}(t,t) \Gamma_{1} \phi(t,{\cal F}(p_1)) ~\tau_{u}(t,t') \Gamma_0' \phi(t',{\cal F}_0'(-p))]\FC\nonumber \\
&\langle {\cal O}^{\bar qq}(t)|{\cal O}_{1/2 }^{MM\dagger}(t')\rangle =   C_{\bar qq}~  \sqrt{\tfrac{3}{2}}~  Tr[\tau_{s}(t',t)\Gamma_0 \phi(t,{\cal F}_0(p))\nonumber \\
& \times ~\tau_{u}(t,t') \Gamma_{1}' \phi(t',{\cal F}(-p_1'))~ \tau_{u}(t',t') \Gamma_2' \phi(t',{\cal F}(-p_2'))]\FD\nonumber
\end{align}
These contractions within the distillation method agree with the contractions within the conventional method derived in Ref. \cite{Nagata:2008wk}. 

The $MM\to MM$ contractions involve three different types. The first two are connected (Fig.~\ref{fig:contractions}a) and can be handled with conventional methods, while the third one (Fig.~\ref{fig:contractions}c) involves backtracking loops with  $\tau(t,t)$ and $\tau(t',t')$, so it needs ``all-to-all'' methods like for example distillation   
\begin{align}
 C^{direct}&(t,t')=\\
Tr[&\tau_{s}(t',t) \Gamma_2 \phi(t,{\cal F}(p_2)) ~\tau_{u}(t,t') \Gamma_2' \phi(t',{\cal F}(-p_2')) ]~\nonumber\\
\times ~  Tr[&\tau_{u}(t',t) \Gamma_{1} \phi(t,{\cal F}(p_1))~ \tau_{u}(t,t') \Gamma_{1}' \phi(t',{\cal F}(- p_1'))]\FC\nonumber\\
C^{crossed}&(t,t')=\nonumber\\
 Tr[&\tau_{s}(t',t) \Gamma_2 \phi(t,{\cal F}(p_2))~ \tau_{u}(t,t') \Gamma_{1}' \phi(t',{\cal F}(- p_1')) ~\nonumber\\
\times ~ &\tau_{u}(t',t) \Gamma_{1} \phi(t,{\cal F}(p_1))~ \tau_{u}(t,t') \Gamma_2' \phi(t',{\cal F}(-p_2'))]\FC\nonumber\\
C^{box}&(t,t')=\nonumber\\
 Tr[&\tau_{s}(t',t) \Gamma_2 \phi(t,{\cal F}(p_2)) ~\tau_{u}(t,t) \Gamma_{1} \phi(t,{\cal F}(p_1))~\nonumber\\
\times ~ & \tau_{u}(t,t') \Gamma_{1}' \phi(t',{\cal F}(-p_1'))~ \tau_{u}(t',t') \Gamma_2' \phi(t',{\cal F}(-p_2'))]~.\nonumber
\end{align}
We compute correlation matrices  $C_{jk}(t,t')$ for all initial time slices $t'$ and all final times slices $t$; then we average over $t'$ at fixed $t-t'$. 

\end{appendix}


\bibliographystyle{h-physrev4}
\bibliography{Lgt,interim}

\end{document}